\documentclass[preprint]{aastex}

\usepackage{textcomp}

\newcommand{\s}{\scriptscriptstyle}

\begin{document}

\title{On the Origin of Pluto's Small Satellites by Resonant Transport}
\author{W.~H.~Cheng$^{\rm a}$, S.~J.~Peale$^{\rm b}$, Man Hoi Lee$^{\rm a,c}$}
\affil{$^{\rm a}$Department of Earth Sciences, The University of Hong Kong,
  Pokfulam Road, Hong Kong}
\affil{$^{\rm b}$Department of Physics, University of California,
  Santa Barbara, CA 93106, United States}
\affil{$^{\rm c}$Department of Physics, The University of Hong Kong,
  Pokfulam Road, Hong Kong}

\begin{abstract}
The orbits of Pluto's four small satellites (Styx, Nix, Kerberos, and
Hydra) are nearly circular and coplanar with the orbit of the large
satellite Charon, with orbital periods nearly in the ratios 3:1, 4:1,
5:1, and 6:1 with Charon's orbital period.
These properties suggest that the small satellites were created during
the same impact event that placed Charon in orbit and had been pushed
to their current positions by being locked in mean-motion resonances
with Charon as Charon's orbit was expanded by tidal interactions with
Pluto.
Using the Pluto-Charon tidal evolution models developed by
\cite{Cheng14}, we show that stable capture and transport of a test
particle in multiple resonances at the same mean-motion
commensurability is possible at the 5:1, 6:1, and 7:1
commensurabilities, if Pluto's zonal harmonic $J_{2{\s P}} = 0$.
However, the test particle has significant orbital eccentricity at the
end of the tidal evolution of Pluto-Charon in almost all cases, and
there are no stable captures and transports at the 3:1 and 4:1
commensurabilities.
Furthermore, a non-zero hydrostatic value of $J_{2{\s P}}$ destroys
the conditions necessary for multiple resonance migration.
Simulations with finite but minimal masses of Nix and Hydra also fail
to yield any survivors.
We conclude that the placing of the small satellites at their current
orbital positions by resonant transport is extremely unlikely.
\end{abstract}

\section{INTRODUCTION}
\label{intro}

Pluto has five known satellites Charon, Styx, Nix, Kerberos, and
Hydra, in the order of distance from Pluto.
Charon was discovered in 1978 \citep{Christy78}, Nix and Hydra in 2005
\citep{Weaver06}, and Kerberos and Styx in 2011 and 2012
\citep{Showalter11,Showalter12}.
Charon is much larger than the other four satellites, and Nix and
Hydra are in turn larger than Kerberos and Styx.\footnote{
Kerberos and Styx are 10\% and 4\% as bright as Nix, respectively
\citep{Showalter11,Showalter12}, which means that they are $\sim
30$ and $100$ times less massive than Nix, if they have similar
densities and albedos.
\label{footnote:masses}
}
The system has nearly coplanar orbital geometry.
The orbital periods are nearly in the ratios of 1:3:4:5:6, but
sufficiently distinct from integer ratios relative to Charon that the
small satellites are not in mean-motion resonances (MMR) with Charon
(e.g., \citealt{Buie13}).
The orbits of the four small satellites are significantly
non-Keplerian because of the large Charon-Pluto mass ratio ($q =
0.1165$) and, at least in the case of Nix and Hydra, because of the
additional effects of their proximity to the 3:2 mean-motion
commensurability \citep{Lee06}.
Orbits too close to Charon are unstable, and Styx is located near the
inner edge of the stable region \citep{Stern94,Nagy06}.
Orbits can also be destabilized by Nix and Hydra, and Kerberos is
located in the only stable region between Nix and Hydra
\citep{PiresdosSantos11,Youdin12}.

The most likely scenario for the formation of Charon is a glancing
impact where the impactor came off nearly intact in an eccentric
orbit with semimajor axis near $4R_{\s P}$ (where $R_{\s P}$ is the
radius of Pluto), with Pluto spinning rapidly,
consistent with the current angular momentum of
the system \citep{Canup05}. If the small satellites formed
simultaneously, they must end up at orbital radii beyond the 3:1 to
6:1 mean-motion commensurabilities if they are to be captured and
transported in resonances at these commensurabilities as Charon's
orbit tidally expanded to its current semimajor axis of $17R_{\s P}$.

For coplanar orbits, the lowest order terms in the disturbing function
that can be resonant at the $m+1$:1 mean-motion commensurability
exterior to Charon are the following \citep{Murray99}:
\begin{equation}
\Phi_m = \frac{GM_{\s C}}{a}
         \sum_{l=0}^{m} f_{m,l}(\alpha) e^{m-l} e_C^{l} \cos \phi_{m,l} ,
\label{eq:resterms}
\end{equation}
where the resonance variables are
\begin{equation}
\phi_{m,l} = (m+1) \lambda - \lambda_{\s C} - (m-l) \varpi -
            l \varpi_{\s C} ,
\label{eq:resvar}
\end{equation}
$G$ is the gravitational constant, $M_{\s C}$ is the mass of Charon, $a$,
$e$, $\lambda$, and $\varpi$ are the orbital semimajor axis,
eccentricity, mean longitude, and longitude of periapse of the small
satellite, and the orbital elements with the subscript C are those of
Charon.
The quantities $f_{m,l}$ are functions of $\alpha = a_{\s C}/a$, the
Laplace coefficients, and their derivatives with respect to $\alpha$,
whose forms are given in Appendix B of \cite{Murray99} for $m \le 4$
using a different subscript notation for $f$.

The motion within each individual resonance term can be represented by
a pendulum equation, where the coefficient of the cosine term in the
disturbing function appears in the restoring acceleration
(e.g., \citealp{Murray99}). Any resonance variable can then be either
circulating or librating about a constant value, but the latter can
occur only if the eccentricity factors are non-zero. A problem with
pushing the small satellites out in MMR with Charon is that the
eccentricity of the small satellite rapidly increases to the point
where the system becomes unstable if the resonance is one of the $l
\neq m$ Lindblad resonances containing powers of $e$ in the
coefficient of the restoring torque (e.g., \citealp{Ward06}).

One notes, however, that there is one resonance term with $l = m$ in
each set (Eq.~[\ref{eq:resterms}]) that only contains powers of
$e_{\s C}$ in the coefficient.
These are called corotation resonances, because the perturbed
satellite librates in a potential field that co-rotates with Charon. 
\citet{Ward06} took advantage of this and 
proposed to transport Nix and Hydra in corotation resonances
at the 4:1 and 6:1 commensurabilities, where the eccentricity
does not grow, and where the resonances are destroyed as
$e_{\s C}\rightarrow 0$ as the current dual synchronous equilibrium
configuration of Pluto-Charon
is approached. But \citet{Lithwick08a} showed that the necessary
conditions for migrating each satellite in a corotation resonance
cannot be satisfied for Nix and Hydra simultaneously
under the same assumptions used by \citeauthor{Ward06}
(i.e., zero gravitational harmonic coefficient $J_{2{\s P}}$ of
Pluto,\footnote{
The gravitational harmonic coefficients $J_2$ and $C_{22}$ of a body
of mass $M$ and radius $R$ are given by $J_2 = [\mathcal{C} -
(\mathcal{A} + \mathcal{B})/2]/(M R^2)$ and $C_{22} = (\mathcal{B} -
\mathcal{A})/(4 M R^2)$, where $\mathcal{A} \le \mathcal{B} \le
\mathcal{C}$ are the principal moments of inertia.
\label{footnote:j2c22}
}
zero mass for Nix and Hydra, and imposed expansion of Charon's orbit.)

Here we investigate the possible transport of the small satellites
locked in multiple MMRs at the same commensurability, as Charon's
orbit is expanded according to conventional tidal models.
\citet[][hereafter paper I]{Cheng14} have studied in detail the tidal
evolution of Pluto-Charon in two tidal models with the frequency $f$
dependence of the dissipation function $Q\propto 1/f$ or $Q =$
constant, including dissipation in both Pluto and Charon and the
possibility of permanent deviations from axisymmetry in both bodies.
We showed in paper I that the inclusion of the effects of the
gravitational harmonic coefficient $C_{22{\s C}}$ of Charon (see
footnote \ref{footnote:j2c22}) for both
rotational and orbital motions can have a profound effect on the
orbital evolution when Charon is captured into spin-orbit resonances
during the evolution.\footnote{
A brief summary of some of the material in both paper I and this paper
appeared in an extended abstract by \cite{Peale11}.}

In Section 2 we demonstrate simultaneous capture in all or several of
the resonances at the 5:1, 6:1, and 7:1 commensurabilities under the
same assumptions used by \cite{Ward06} and \cite{Lithwick08a}, where
$J_{2{\s P}}=0$ and the small satellites are treated as test
particles, but with a more realistic tidal expansion of the orbit of
Charon. The eccentricity of Charon's orbit remains  
non-zero throughout most of the tidal evolution to the current
configuration by a judicious choice of the ratio of tidal dissipation
in Charon to that in Pluto (paper I), thereby maintaining
the stability of the resonances. In some cases, the eccentricity of the
test particle does not grow during the transport, which would occur in
single Lindblad resonance capture. Although this seems like a possible
solution to the eccentricity growth problem and the unlikely selection
of only corotation resonances as the small satellites encounter the
respective mean-motion commensurabilities with Charon, the test
particle's final eccentricity is still too large, and we are unable
to demonstrate a similar multiple-resonance migration
at the 3:1 and 4:1 commensurabilities.
In Section 3 we show how a non-zero $J_{2{\s P}}$
destroys the migration stability we find for test particles for a
large ensemble of plausible initial conditions. Finally, the
Pluto system with non-zero, but minimal masses for Nix and Hydra is
integrated in Section 4 for plausible ranges of parameters in a search
for an overlooked configuration that could lead to resonant transport
of Nix and Hydra. None are found.
We summarize and discuss the significance of these results in Section
5.

\section{MULTI-RESONANCE CAPTURE AND TRANSPORT}
\label{multi}

We study the behavior of the debris from the collisional capture of
Charon by adding test particles to the simulations of the tidal
evolution of Pluto-Charon presented in paper I.
We use the Wisdom-Holman \citeyearpar{Wisdom91} integrator in the
SWIFT\footnote{See http://www.boulder.swri.edu/$\sim$hal/swift.html.}
package \citep{Levison94}, modified to simulate the tidal, rotational
and axial-asymmetry effects for Pluto and Charon (see paper I for
details).
The Jacobi coordinates are adopted for the test particles, following
\citet{Lee06}.
We use a slightly different division of the Hamiltonian into the
Keplerian and perturbation parts, which is designed for integrations
with comparable masses (such as Pluto-Charon) in a hierarchical system
(\citealp{Lee03}; see also \citealp{Beust03}).
All external perturbations from the Sun and the other planets are
negligibly small.

We assume a typical outcome for the collisional capture of Charon,
where Charon emerges with non-zero orbital eccentricity
$e_{\s C}$ and semimajor axis $a_{\s C} = 4R_{\s P}$
\citep{Ward06}.
Both Pluto and Charon would be spinning after the collision (with both
spin axes assumed to be perpendicular to the orbit plane), where the
total angular momentum is that  of the current system.
Charon's spin contributes relatively little to the total angular
momentum, and it evolves quickly to the asymptotic tidal spin rate
(where the tidal torque averaged over an orbit vanishes),
unless it is captured temporarily into spin-orbit resonances due to
non-zero $C_{22{\s C}}$.
For the simulations in this section, we assume $J_{2{\s P}} = 0$.

For the tidal model where the tidal distortion of a body responds to
the perturbing body a short time $\Delta t$ in the past, constant
$\Delta t$ leads to  $Q\propto 1/f$, so we call the $Q\propto 1/f$
model the constant $\Delta t$ model.
As shown in paper I, the eccentricity $e_{\s C}$ of Charon's orbit can
be roughly constant during most of the tidal evolution with the
appropriate ratio of tidal dissipation in Charon to that in Pluto
characterized by
\begin{equation}
A_{\Delta t} = \frac{k_{2C}}{k_{2P}} \frac{\Delta t_{\s C}}{\Delta t_{\s P}}
\left( \frac{M_{\s P}}{M_{\s C}} \right)^2 \left( \frac{R_{\s C}}{R_{\s P}} \right)^5
\approx \frac{\mu_{\s P}}{\mu_{\s C}} \frac{\Delta t_{\s
C}}{\Delta t_{\s P}} \frac{R_{\s C}}{R_{\s P}} ,
\label{eq:a1}
\end{equation}
where the subscripts $P$ and $C$ denote Pluto and Charon,
respectively, $M_i$ is the mass, and $R_i$ is the radius.
The second degree potential Love number
\begin{equation}
k_{2i} = {3/2 \over 1+19\mu_i/(2\rho_i g_i R_i)}
\end{equation}
for an incompressible homogeneous sphere of radius $R_i$, rigidity
$\mu_i$, density $\rho_i$, and surface gravity $g_i$.
We have used the approximation $k_{2i} \approx 3 \rho_i g_i R_i/
(19 \mu_i)$, valid for small solid body, in the last form of
Eq.~(\ref{eq:a1}).

We perform integrations with a range of eccentricity evolution of
Charon by varying initial $e_{\s C}$ and $A_{\Delta t}$.
Charon on a highly eccentric orbit would lead to close encounters with
the test particles and destabilize the resonances.
Hence we select the range of initial $e_{\s C}$ from 0.05 to 0.3, in
steps of 0.05.
In these runs, we use $k_{2{\s P}} = 0.058$ and $\Delta t_{\s P}=600$
seconds, same as that in paper I.
The values of $A_{\Delta t}$ are chosen to center at the value where $e_{\s C}$
is roughly constant during most of the evolution: $A_{\Delta t} = 9$,
$10$, and $11$.
Smaller (larger) $A_{\Delta t}$ would result in $e_{\s C}$ increasing
(decreasing) throughout most of the evolution.
We assume $C_{22i} = 10^{-5}$.
For the constant $\Delta t$ model, non-zero $C_{22i}$ does not change
the tidal evolution with initial $e_{\s C} \le 0.25$.
For initial $e_{\s C} = 0.3$, Charon can be captured into the 3:2
spin-orbit resonance, but the capture has only small effects on the
evolution of $a_{\s C}$ and $e_{\s C}$ (see paper I).
So we have not repeated the runs with $C_{22i} = 0$.

\begin{figure}[t]
\centering
\epsscale{1.0}
\plottwo{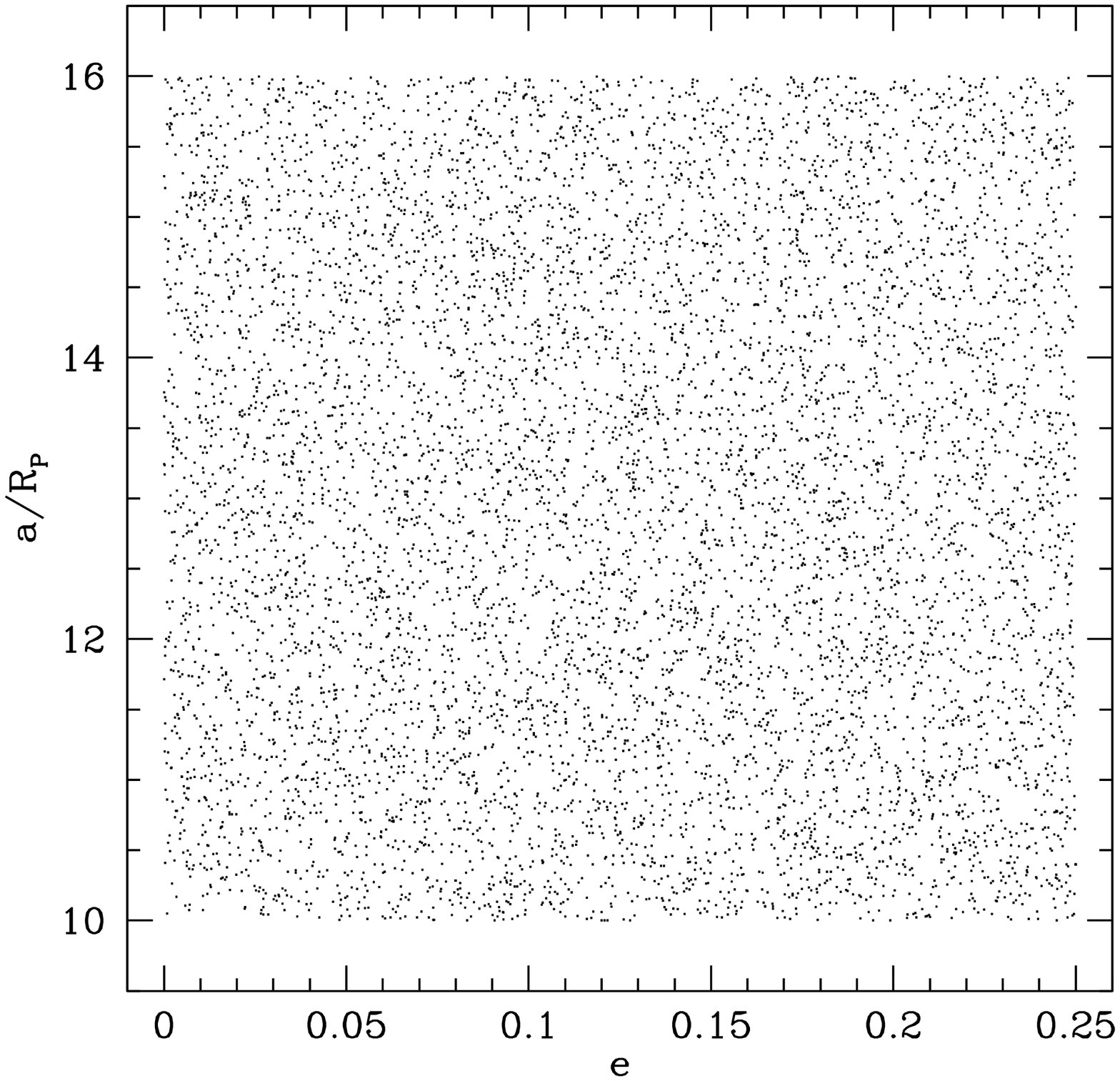}{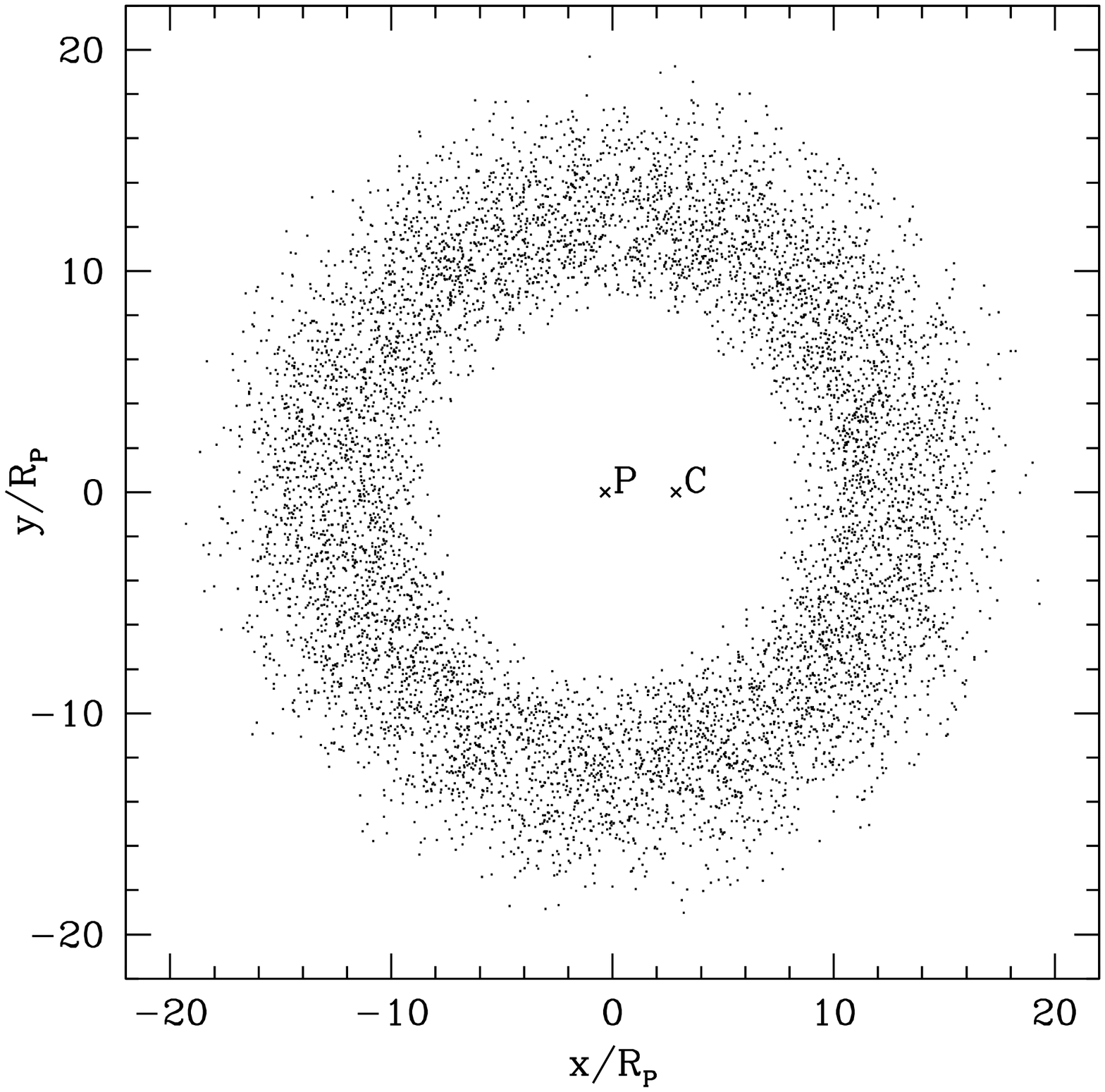}
\caption{
Initial distribution of the 8000 random test particles in $(e,a)$
plane (left panel) and Cartesian $(x,y)$ plane (right panel).
The initial positions of Pluto and Charon with $e_{\s C}=0.2$ and
Charon at periapse are shown as well in the right panel.}
\label{ics}
\end{figure}

For each combination of initial $e_{\s C}$ and $A_{\Delta t}$, 8000 test
particles in orbits coplanar with that of Pluto-Charon are distributed
with random initial conditions.
The semimajor axis $a$ spans from $10R_{\s P}$ to $16R_{\s P}$, which
corresponds to just within 4:1 up to 8:1 mean-motion ratio.
The formation of a massive satellite together with a debris disk
up to this extent by a single giant impact has been confirmed
by high resolution SPH simulations \citep{Canup11}.
The eccentricity $e$ is between 0 and 0.25, and the longitude of
periapse $\varpi$ and mean anomaly $\mathcal{M}$ are both randomly
chosen.
The initial $(e,a)$ and $(x,y)$ distributions of the test particles
are shown in Fig.~\ref{ics}.
The simulations are first run for $10^{10}$ seconds ($\approx 300$ years).
We refer to this as the first stage of our integration.

\begin{figure}[t]
\centering
\epsscale{1.0}
\plottwo{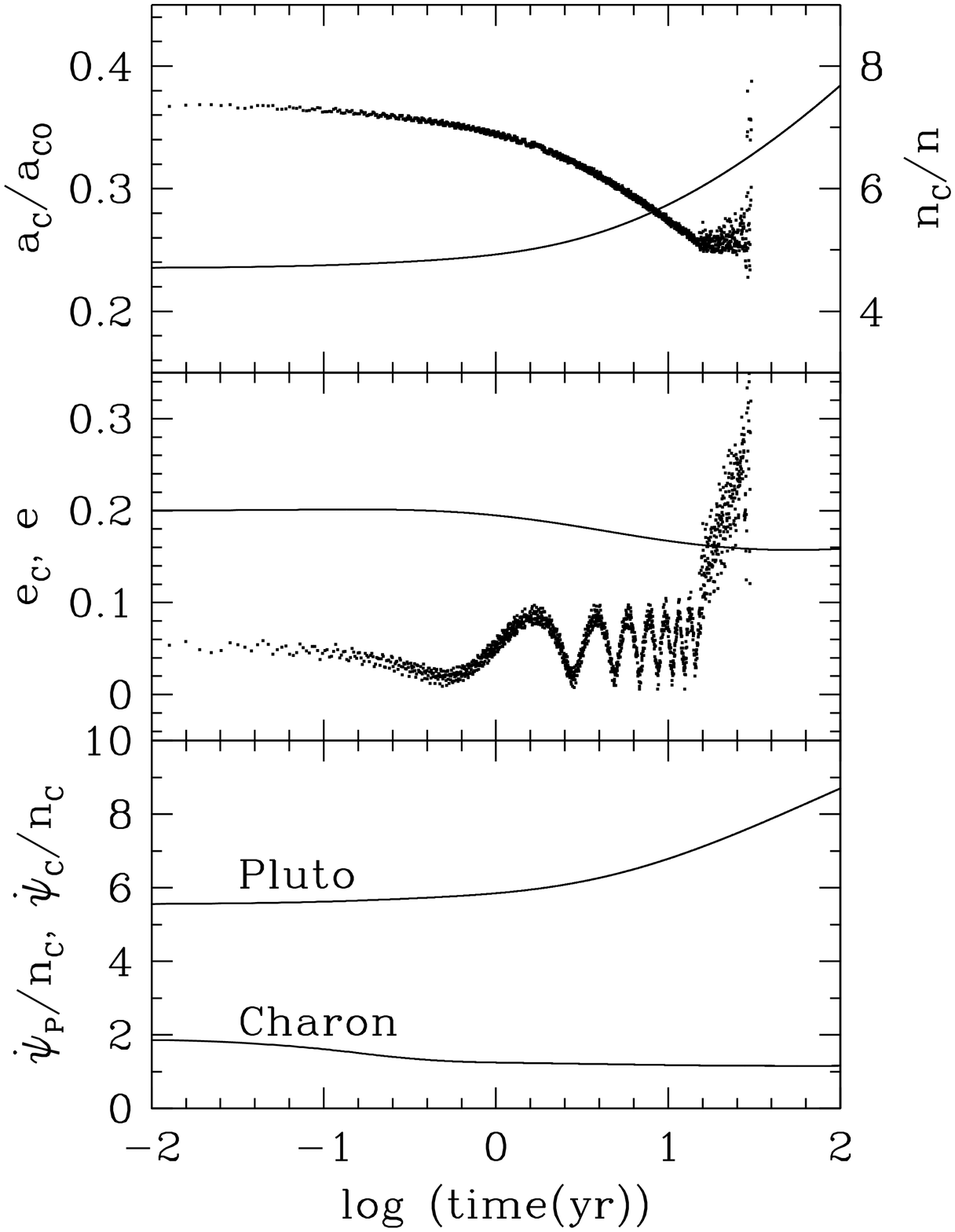}{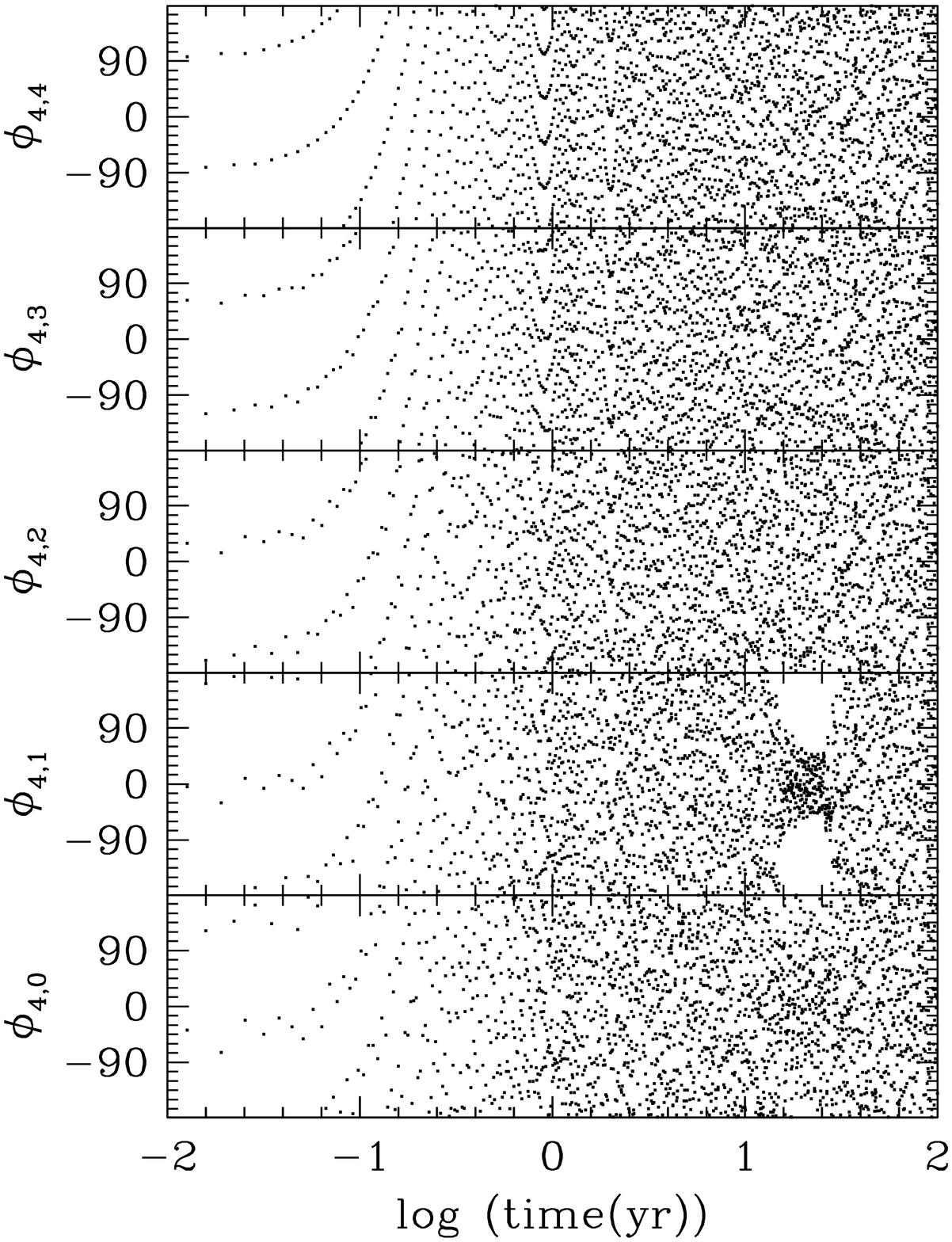}
\caption{
Example of a test particle captured into a single 5:1 Lindblad
resonance with Charon in the constant $\Delta t$ model.
The panels show the evolution of Charon's orbital semimajor axis
$a_{\s C}$ (in units of its current semimajor axis $a_{{\s C}0}$) and
eccentricity $e_{\s C}$ and the spin angular velocities
${\dot \psi}_i$ of Pluto and Charon (in units of Charon's mean motion
$n_{\s C}$) as lines, and the evolution of $e$, the ratio of mean
motion $n_{\s C}/n$, and the resonance variables $\phi_{4,l}$ of the
test particle as dots.
$e$ increases when $\phi_{4,1}$ is librating, and the test particle
becomes unstable.}
\label{sinlin}
\end{figure}

Most of the test particles ($\approx 99\%$) become unstable and are
removed
during the first stage of our integration.
Test particles are removed if they have close encounter with either
Pluto or Charon or get too far away ($\gtrsim 100 R_P$).
About half of them are removed either very quickly without getting
into any identifiable resonance with Charon, or as soon as they
encounter one of the commensurabilities with Charon.
A significant fraction of the latter are caught into a single
Lindblad resonance between 4:1 and 8:1, which leads to
excitation of their eccentricities.
They become unstable and are removed eventually, as predicted by
\citet{Ward06}.
We present one of these in Fig.~\ref{sinlin} as an example.
The figure shows the evolution of $a_{\s C}$, $e_{\s C}$, and the spin
angular velocities $\dot{\psi}_i$ as lines, and the evolution of the
orbital eccentricity $e$, mean motion ratio $n_{\s C}/n$, and resonance
variables $\phi_{4,l}$ (Eq.~[\ref{eq:resvar}] with $m = 4$) of the
test particle as dots.
Although Pluto's spin is decreasing, the orbital mean motion is
decreasing even faster so $\dot{\psi}_{\s P}/n_{\s C}$ is rising.
The orbital elements of the test particle are the osculating Keplerian
ones in Jacobi coordinates.
Before being captured, the period ratio decreases as $a_{\s C}$ increases.
After that, the ratio fluctuates around the value of that
commensurability (5:1). 
Since Charon is massive and its orbit is eccentric, the orbit of the
test particle is significantly non-Keplerian, as described by
\citet{Lee06} and \citet{Leung13}.
The osculating Keplerian eccentricity $e$ undergoes short period,
small amplitude oscillations forced by the non-axisymmetric components
of the potential of Pluto-Charon,
and longer period oscillations from the superposition of the epicyclic
motion and the forced eccentricity oscillation.
$e$ starts to grow as soon as the test particle is caught into a
single $m = 4$, $l = 1$ Lindblad resonance, and the test particle
becomes unstable and is removed when $e$ becomes too high.

We have several tens of survivors in each simulation when we first
stop the integrations, except for initial $e_{\s C}=0.05$,
where no particle survives. All of the survivors have been
caught in multiple resonances at each mean-motion commensurability,
which allow stable expansion of the test particle's orbit as Charon's
orbit expands, but none of them are in 3:1 or 4:1.
(Recall that Styx, Nix, Kerberos, and Hydra are near 3:1, 4:1, 5:1 and
6:1, respectively.)
There were no captures into just the corotation resonances
with subsequent stable expansion of the test particle orbit,
and no stable expansions without simultaneous libration
in multiple resonances at the same commensurability.
In Table \ref{dtstat}, we list the statistics of test particles
remaining in 5:1, 6:1, and 7:1 resonances when we first stop the
integrations.
In a few cases, we find particles that have already escaped from
multiple resonances at a mean-motion commensurability but not yet
removed at the end of the first stage of our integration.

\begin{deluxetable}{lcccccccccccc}
\tablecolumns{10}
\tablewidth{0pt}
\tablecaption{Fraction of captured test particles,
in unit of permille (\textperthousand),
after $10^{10}$ seconds in the constant $\Delta t$ model
with $C_{22i} = 10^{-5}$
\label{dtstat}}
\tablehead{
\colhead{} & \colhead{} & \multicolumn{3}{c}{$A_{\Delta t} = 9$} & \colhead{} &
\multicolumn{3}{c}{$A_{\Delta t} = 10$} & \colhead{} & \multicolumn{3}{c}{$A_{\Delta t} = 11$} \\
\cline{3-5} \cline{7-9} \cline{11-13}\\
\colhead{Initial $e_{\s C}$} & & 5:1 & 6:1 & 7:1 & & 5:1 & 6:1 & 7:1 &
& 5:1 & 6:1 & 7:1
}
\startdata
0.10 & & 7.75 & 2.50 & 0 & & 9.63 & 0.63 & 0 & & 6.00 & 0 & 0 \\
0.15 & & 1.00 & 11.00 & 4.13 & & 4.25 & 9.38 & 3.00 & & 5.75 & 10.25 & 2.50 \\
0.20 & & 0.88 & 6.88 & 5.00 & & 1.25 & 8.75 & 5.38 & & 3.13 & 11.13 & 4.89 \\
0.25 & & 0.25 & 4.38 & 6.13 & & 0.75 & 4.00 & 6.00 & & 0.50 & 7.38 & 5.88 \\
0.30 & & 0.25 & 4.13 & 6.13 & & 0.50 & 3.25 & 7.38 & & 0.25 & 4.88 & 6.75 \\
\enddata
\end{deluxetable}

\begin{deluxetable}{lcccccccccccc}
\tablecolumns{10}
\tablewidth{0pt}
\tablecaption{Fraction of survivors, in unit of permille (\textperthousand),
to the end of the tidal evolution of Pluto-Charon
in the constant $\Delta t$ model with $C_{22i} = 10^{-5}$.
\label{dtstatend}}
\tablehead{
\colhead{} & \colhead{} & \multicolumn{3}{c}{$A_{\Delta t} = 9$} & \colhead{} &
\multicolumn{3}{c}{$A_{\Delta t} = 10$} & \colhead{} & \multicolumn{3}{c}{$A_{\Delta t} = 11$} \\
\cline{3-5} \cline{7-9} \cline{11-13}\\
\colhead{Initial $e_{\s C}$} & & 5:1 & 6:1 & 7:1 & & 5:1 & 6:1 & 7:1 &
& 5:1 & 6:1 & 7:1
}
\startdata
0.1 & & 0 & 0.25 & 0 & & \nodata & \nodata & \nodata & & 0 & 0 & 0 \\
0.2 & & \nodata & \nodata & \nodata & & 0 & 0.63 & 1.00 & & \nodata & \nodata & \nodata \\
0.3 & & 0 & 0.13 & 0 & & \nodata & \nodata & \nodata & & 0 & 1.88 & 1.13 \\
\enddata
\tablecomments{
The blank spaces correspond to values of $e_{\s C}$ and $A_{\Delta t}$
for which the integrations were not continued beyond $10^{10}$ seconds.
}
\end{deluxetable}

We observe a general trend that the fraction of captures into multiple
5:1 resonances decreases when initial $e_{\s C}$ increases, and the
opposite happens for 7:1 resonances.
This trend and the result that no particle survives for initial
$e_{\s C}=0.05$ can be understood qualitatively.
If initial $e_{\s C}$ is too small, the resonance widths are too
narrow for simultaneous libration in multiple resonances.
For 6:1, as we increase initial $e_{\s C}$, simultaneous libration in
multiple resonances first becomes more stable as the resonance widths
increase with $e_{\s C}$, but then becomes less stable when Charon
comes too close at apopase and the perturbation from Charon becomes
too strong.
For the higher-order resonance 7:1 further from Charon, the fraction
of captured test particles continues to increase up to initial
$e_{\s C} = 0.3$.
But for the lower-order resonance 5:1 closer to Charon, the fraction
of captured test particles already decreases beyond initial
$e_{\s C} = 0.1$.
From this qualitative understanding of the trends in the numerical
results, we believe we have focused on the suitable combinations of
initial $e_{\s C}$ and $A_{\Delta t}$ in exploring the multiple
resonance migration scenario.

We continue to integrate the particles that survive after our first
stage of integration to the end of the tidal evolution of Pluto-Charon
for the four corners and the middle point of the $(e_{\s C},
A_{\Delta t})$ grid
in Table \ref{dtstat}.
Most of the test particles escape from resonances and are removed,
with the majority removed during the final decrease in $e_{\s C}$ near
the end of the tidal evolution.
The final survived fractions of the initial 8000 test particles are
summarized in Table \ref{dtstatend}.

\begin{figure}[t]
\centering
\plottwo{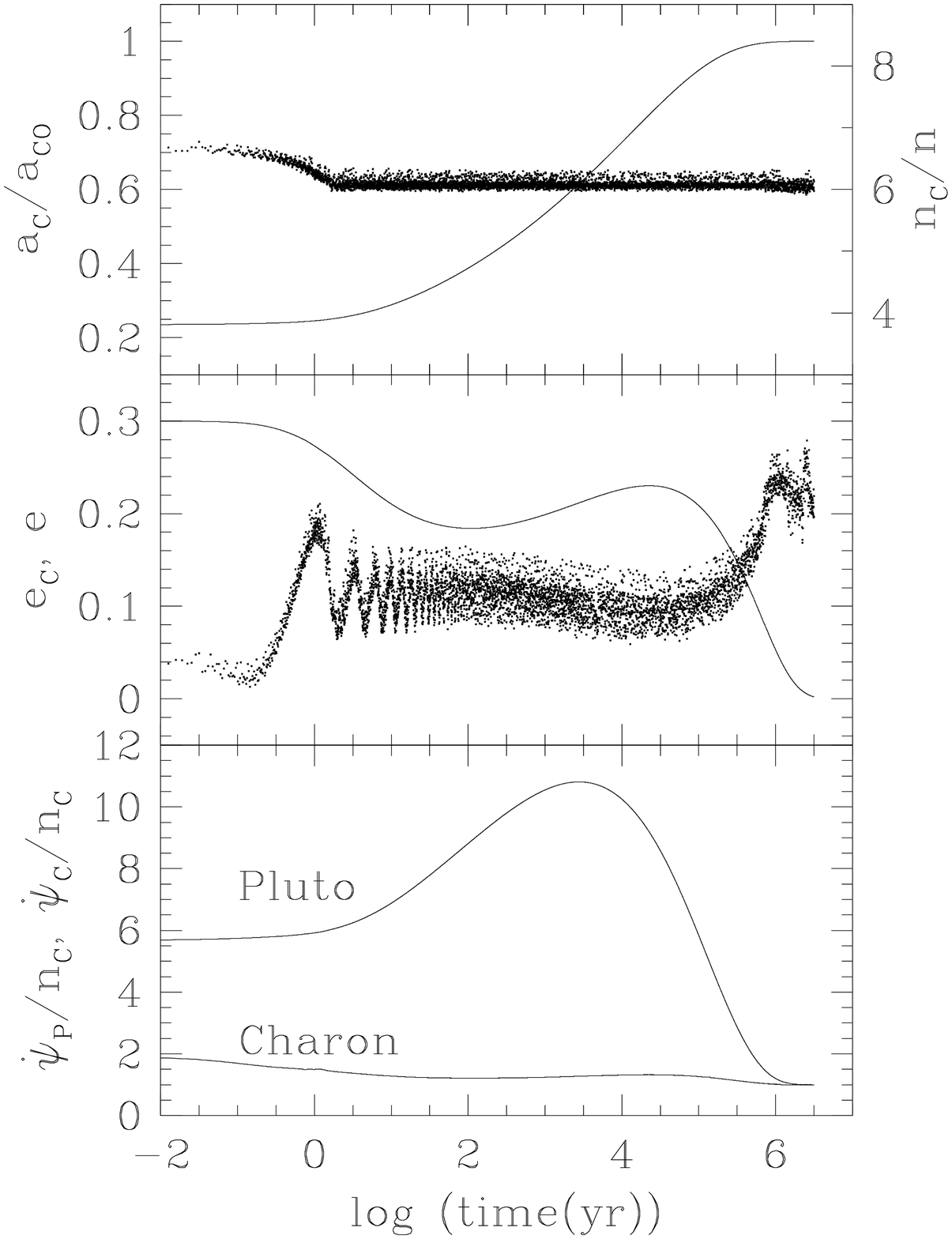}{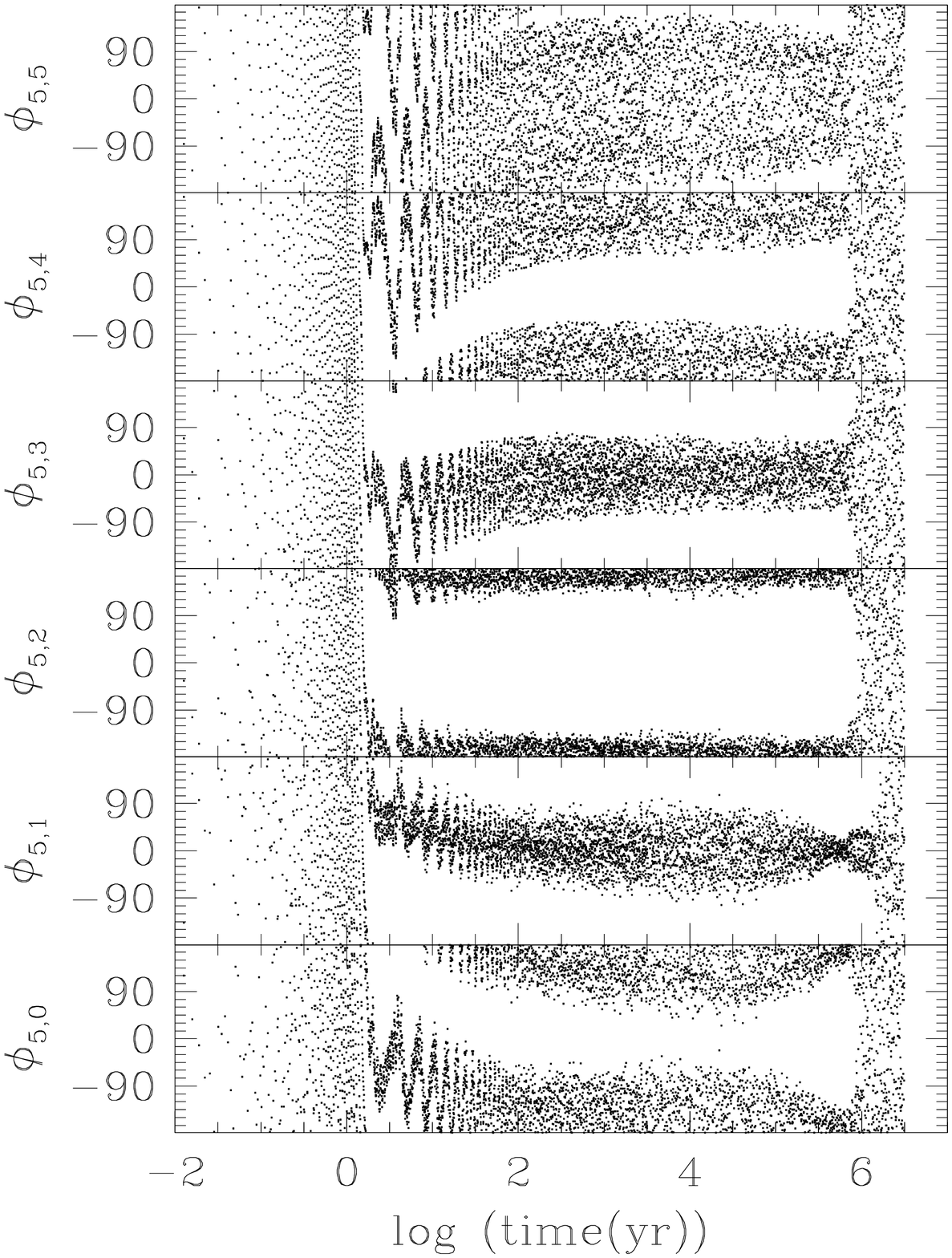}
\caption{Example of a test particle that survives to the end of
the tidal evolution of Pluto-Charon in the constant $\Delta t$ model.
The test particle is captured into simultaneous libration of all six
resonance variables $\phi_{5,l}$ at the 6:1 commensurability with
Charon, with the orbits anti-aligned.
$e$ increases to above $\approx 0.2$ with decreasing $e_{\s C}$ as
Pluto and Charon approach the dual synchronous state.
Initial $e_{\s C}=0.3$ and $A_{\Delta t}=11$.}
\label{egdtleft}
\end{figure}

Fig.~\ref{egdtleft} shows an example of a test particle that survives
to the end of the tidal evolution of Pluto-Charon.
The left panel shows the evolution of the semimajor axes,
eccentricities, and spins as Charon's semimajor axis expands from
$4R_{\s P}$ to its current value of $17R_{\s P}$,
and the right panel shows the evolution of the six resonance variables
$\phi_{5,l}$ at the 6:1 commensurability with Charon.
The test particle is captured into 6:1 resonance, with the
simultaneous libration of all six resonance variables.
The resonance variables alternate between libration about $0^\circ$
and $180^\circ$, and the orbits are anti-aligned with the difference
in the longitudes of periapse $\varpi-\varpi_{\s C}=180^\circ$.
The eccentricity of the test particle $e$ does not exceed 0.2 for most
of the evolution, when the eccentricity of Charon $e_{\s C}$ remains
significant.
As $a_{\s C}$ approaches $17R_{\s P}$ and Charon and Pluto approach
synchronous rotation, the test particle escapes from all six 6:1
resonances, but it stays near the 6:1 commensurability.
However, $e$ increases as $e_{\s C}$ decreases on the approach of
Pluto-Charon to the dual synchronous state.
So even though the test particle is migrated successfully to
the current semimajor axis of Hydra, it does not resemble
the characteristics of Hydra's orbit, because tides at its current
semimajor axis are too weak to damp its eccentricity in the age of the
Solar System \citep{Stern06}.

\begin{deluxetable}{lcccccccccccc}
\tablecolumns{10}
\tablewidth{0pt}
\tablecaption{Fraction of captured test particles, in unit of permille
(\textperthousand), after $3 \times 10^{11}$ seconds
in the constant $Q$ model with $C_{22i}=0$.
\label{qstat}}
\tablehead{
\colhead{} & \colhead{} & \multicolumn{3}{c}{$A_Q= 0.55$ or 1.13} &
\colhead{} & \multicolumn{3}{c}{$A_Q= 0.65$ or 1.14} & \colhead{} &
\multicolumn{3}{c}{$A_Q= 0.75$ or 1.15} \\
\cline{3-5} \cline{7-9} \cline{11-13}\\
\colhead{Initial $e_{\s C}$} & & 5:1 & 6:1 & 7:1 & & 5:1 & 6:1 & 7:1 &
& 5:1 & 6:1 & 7:1
}
\startdata
0.1 & & 1.63 & 4.63 & 1.13 & & 2.88 & 6.25 & 1.25 & & 3.75 & 5.13 & 0.38 \\
0.2 & & 0.38 & 2.88 & 2.75 & & 0.50 & 4.00 & 2.75 & & 0.50 & 3.88 & 3.50 \\
0.3 & & 0.13 & 1.50 & 3.38 & & 0.25 & 1.38 & 3.63 & & 0 & 1.63 & 4.00 \\
\enddata
\tablecomments{
{Larger $A_Q$ for initial $e_{\s C}=0.3$ only.}
}
\end{deluxetable}

\begin{deluxetable}{lcccccccccccc}
\tablecolumns{10}
\tablewidth{0pt}
\tablecaption{Fraction of survivors, in unit of permille (\textperthousand),
to the end of the tidal evolution of Pluto-Charon in the constant $Q$ model
with $C_{22i}=0$.
\label{qstatend}}
\tablehead{
\colhead{} & \colhead{} & \multicolumn{3}{c}{$A_Q= 0.55$ or 1.13} &
\colhead{} & \multicolumn{3}{c}{$A_Q= 0.65$ or 1.14} & \colhead{} &
\multicolumn{3}{c}{$A_Q= 0.75$ or 1.15} \\
\cline{3-5} \cline{7-9} \cline{11-13}\\
\colhead{Initial $e_{\s C}$} & & 5:1 & 6:1 & 7:1 & & 5:1 & 6:1 & 7:1 &
& 5:1 & 6:1 & 7:1
}
\startdata
0.1 & & 0 & 2.13 & 0.63 & & \nodata & \nodata & \nodata & & 2.50 & 0.75 & 0 \\
0.2 & & \nodata & \nodata & \nodata & & 0.13 & 2.75 & 1.50 & & \nodata
& \nodata & \nodata \\
0.3 & & 0 & 0 & 0.50 & & \nodata & \nodata & \nodata & & 0 & 0.75 & 0.75 \\
\enddata
\tablecomments{
{Larger $A_Q$ for initial $e_{\s C}=0.3$ only.
The blank spaces correspond to values of $e_{\s C}$ and $A_Q$ 
for which the integrations were not continued beyond $3 \times
10^{11}$ seconds.}
}
\end{deluxetable}

The second tidal model we consider is the constant $Q$ model.
Details of this model are also given in paper I. The ratio of
dissipation in Charon to that in Pluto for this model is given by 
\begin{equation}
A_Q = \frac{k_{2C}}{k_{2P}} \frac{Q_{\s P}}{Q_{\s C}} \left( \frac{M_{\s P}}{M_{\s C}} \right)^2 \left( \frac{R_{\s C}}{R_{\s P}} \right)^5
\approx \frac{\mu_{\s P}}{\mu_{\s C}} \frac{Q_{\s P}}{Q_{\s C}} \frac{R_{\s C}}{R_{\s P}} .
\label{eq:a2}
\end{equation}
We use $Q_{\s P}=100$, same as that in paper I.
Again, a range of eccentricity evolution of Charon is explored with a
combination of initial $e_{\s C}$ and $A_Q$:
$A_Q = 0.55$, 0.65, and 0.75 for initial $e_{\s C}=0.1$ and $0.2$, and
$A_Q = 1.13$, 1.14, and 1.15 for initial $e_{\s C}=0.3$.
The resulting evolution of Charon's orbit is displayed in paper I.
The initial distribution of test particles is the same as the constant
$\Delta t$ model above.
The first stage of integration ends after $3 \times 10^{11}$ seconds
($\approx 10^4$ years), when the orbit expansion (increase in
$a_{\s C}$) is comparable to that in the constant $\Delta t$ model.

\begin{figure}[t]
\centering
\plottwo{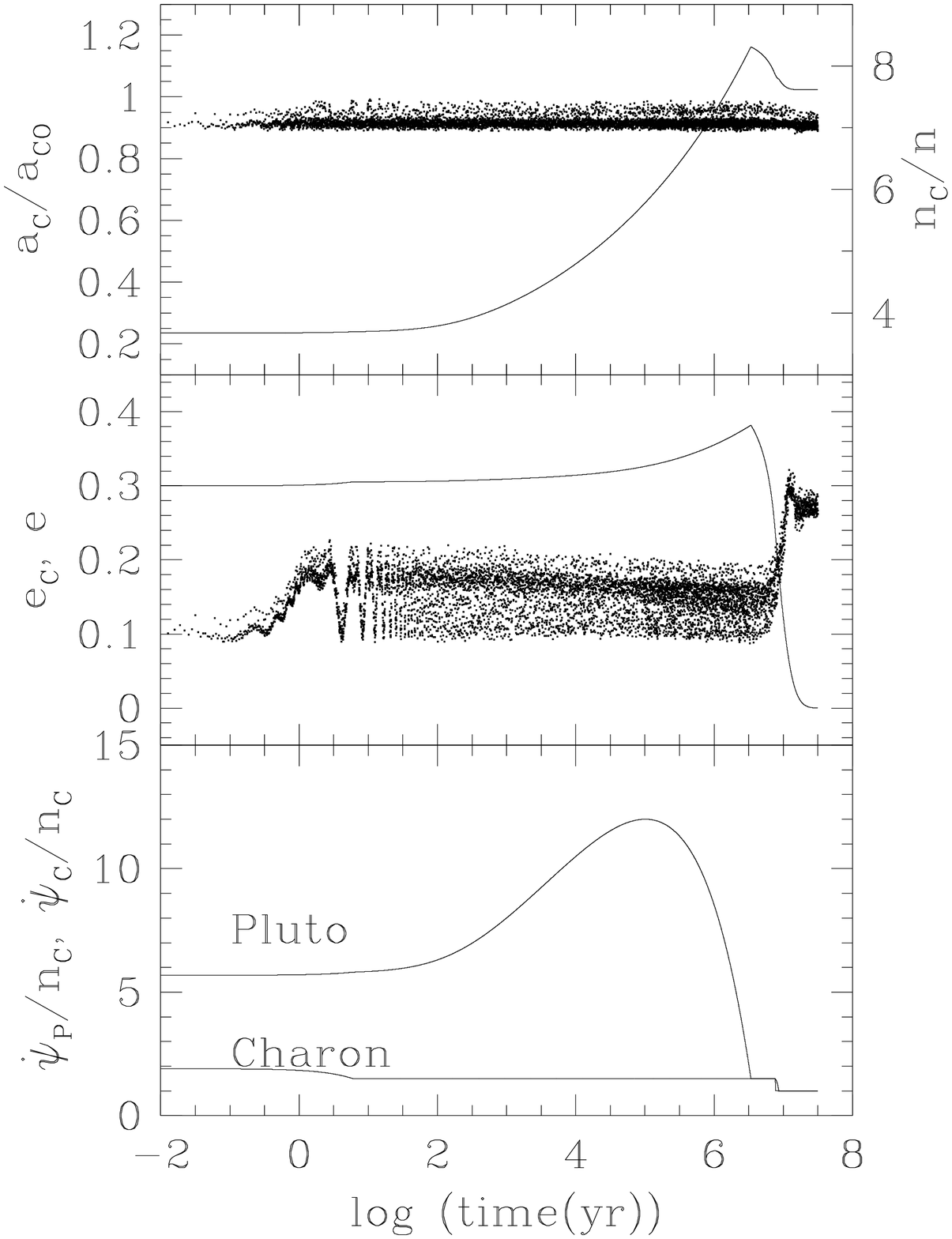}{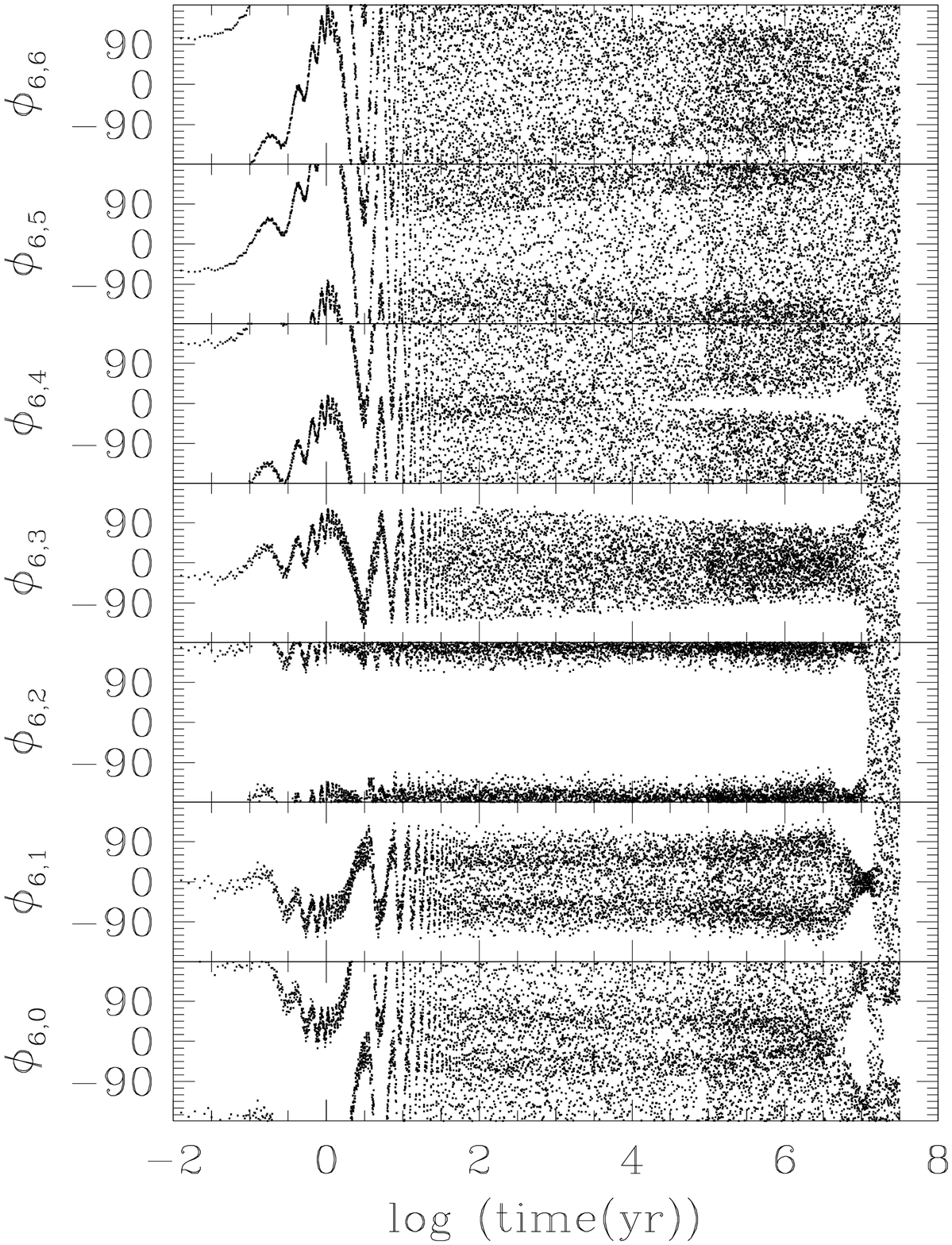}
\caption{
Example of a test particle that survives to the end of the tidal
evolution of Pluto-Charon in the constant $Q$ model.
The test particle is captured and transported in multiple resonances
at the 7:1 commensurability in an anti-aligned configuration with
Charon.
$e$ increases as $e_{\s C}$ decreases on the approach of Pluto-Charon
to the dual synchronous state.
Initial $e_{\s C}=0.3$ and $A_Q=1.13$.}
\label{egqleft}
\end{figure}

We consider first runs with $C_{22i} = 0$.
Table \ref{qstat} shows the statistics of test particles captured into
multiple resonances with stable orbit expansion at the
5:1, 6:1, and 7:1 commensurabilities for the specified values of
$A_Q$ and initial $e_{\s C}$ when the first stage of integration ends.
Like the constant $\Delta t$ model, there is a considerable number
of captures and stable migration at the three commensurabilities, but
none at the 3:1 or 4:1.
Also like the constant $\Delta t$ model, the fraction of captures into
multiple 5:1 resonances decreases with increasing initial $e_{\s C}$,
and the opposite happens for 7:1 resonances.
The results of the continued integrations to the end of the tidal
evolution of Pluto-Charon for the four corners and the middle point of
the ($e_{\s C}, A_Q$) grid are presented in Table \ref{qstatend}.
More than one third of the survivors in the first stage remains.
Despite the evolution timescale in the two tidal models differing by
more than
an order of magnitude, the capture statistics at the end of the first
stage of our integration are within a factor of two, with the constant
$\Delta t$ model more effective for capture (compare Tables
\ref{dtstat} and \ref{qstat}).
On the other hand, the fraction of final survivors at the end of the
tidal evolution in the constant $Q$ model is more than that of
constant $\Delta t$ (compare Tables \ref{dtstatend} and
\ref{qstatend}).

We have also repeated the runs with $C_{22i} = 10^{-5}$ and the same
values of $A_Q$ as those with $C_{22i} = 0$.
For initial $e_{\s C} = 0.1$, non-zero $C_{22i}$ makes no difference
in the tidal evolution of Pluto-Charon, and the test-particle
statistics are comparable to those in Tables \ref{qstat} and
\ref{qstatend}, but fluctuates due to stochastic capture.
However, Charon is captured into 3:2 spin-orbit resonance
for higher initial $e_{\s C}$.
Charon's eccentricity $e_{\s C}$ quickly plummets for initial
$e_{\s C}=0.2$, and rises to $e_{\s C} \ga 0.36$, where the evolution
equations for constant $Q$ are qualitatively inaccurate due to
truncation of the expansion in orbital elements, for initial
$e_{\s C}=0.3$ (see paper I).
None of the test particles survives when $e_{\s C}$ becomes too small
for initial $e_{\s C}=0.2$ or too large for initial $e_{\s C} = 0.3$.

\begin{figure}[t]
\centering
\plottwo{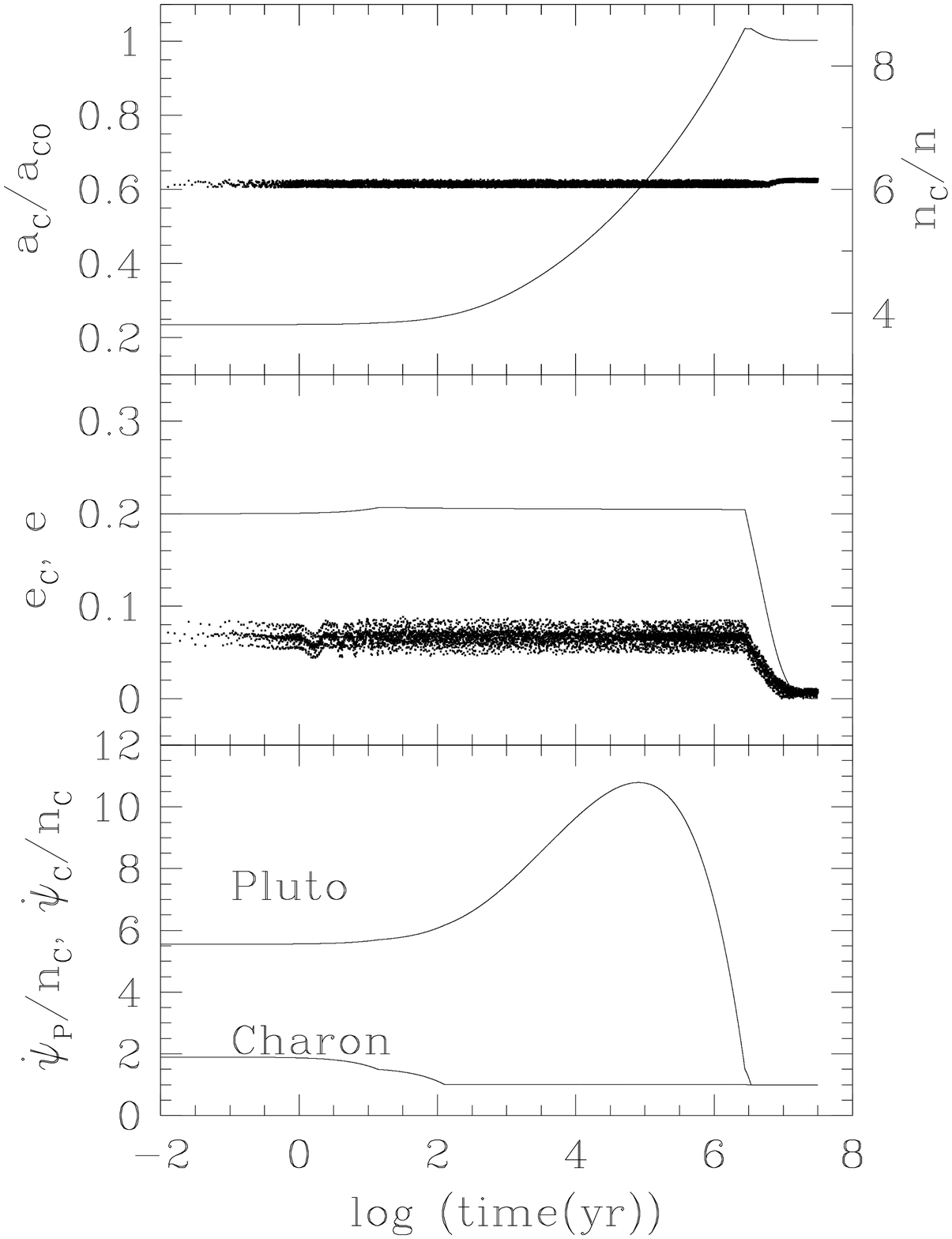}{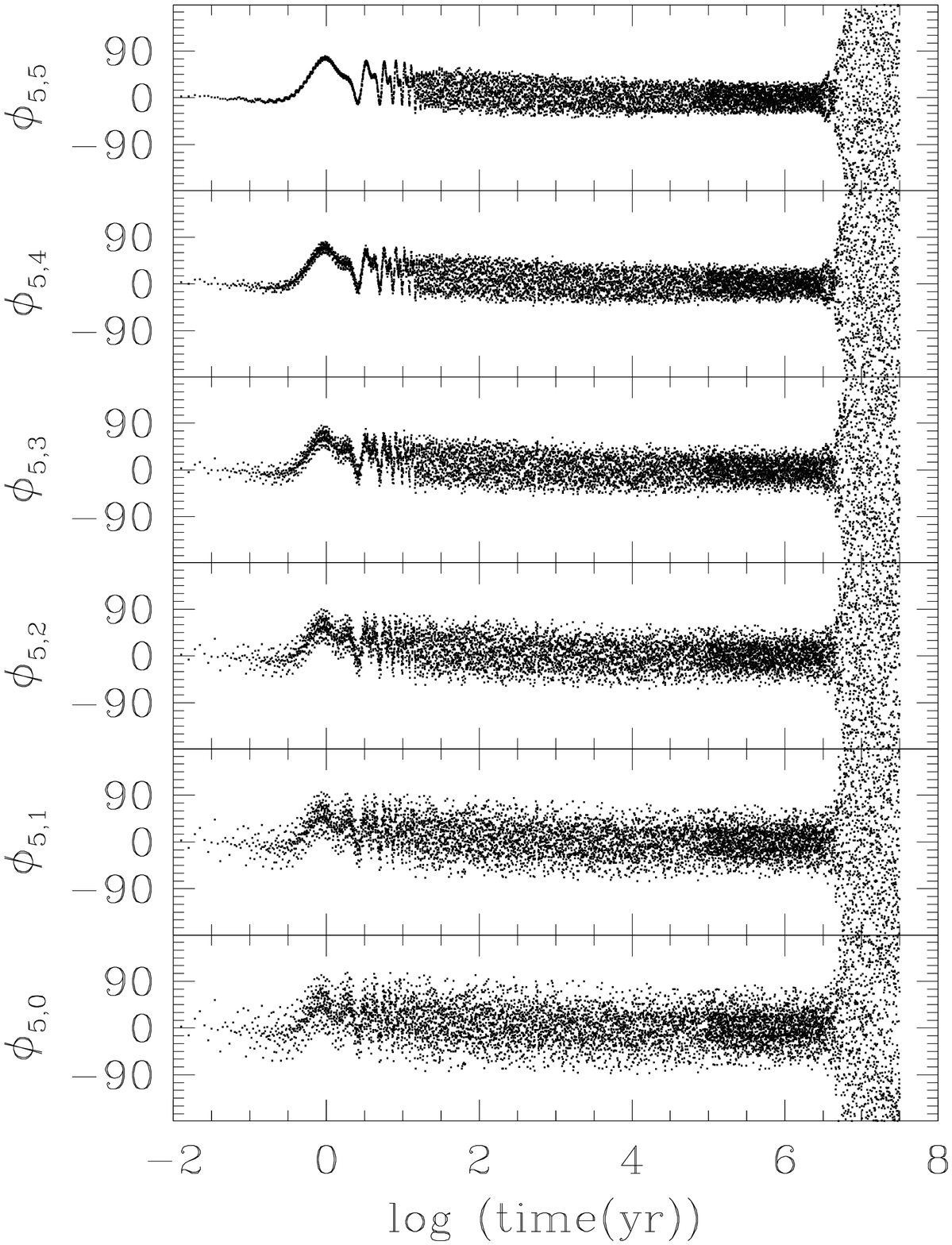}
\caption{
Only case where the test particle is captured and transported in an
aligned configuration with Charon.
The tidal model is constant $Q$, and all 6:1 resonance variables are
librating about $0^\circ$.
Note that in this case $e$ damps to zero along with $e_{\s C}$.
Initial $e_{\s C} = 0.2$ and $A_Q=0.65$.}
\label{onealign}
\end{figure}

An example of a survivor in the constant $Q$ model with $C_{22i} = 0$
is shown in Fig.~\ref{egqleft}.
The test particle is captured and transported in multiple resonances
at the 7:1 commensurability in an anti-aligned configuration with
Charon.
This example is chosen to demonstrate the possibility
of staying within the resonances when $a_{\s C}$ is decreasing.
If $e_{\s C}$ is still large when $a_{\s C}$ reaches the current
value, then by the conservation of angular momentum, $a_{\s C}$ would
overshoot before coming back to the current value when $e_{\s C}$
decays.
At the end of the integration, the test particle is in the $l=0$
resonance only, the coefficient of which does not involve $e_{\s C}$.
We find that around half of the survivors in the two tidal models are
in this situation, and the other half stay near the commensurabilities
but no resonance variable is librating (e.g., Fig.~\ref{egdtleft}).
As in Fig.~\ref{egdtleft}, $e$ increases with decreasing $e_{\s C}$ as
Pluto and Charon approach the dual synchronous state.

Fig.~\ref{onealign} shows a case in the constant $Q$ model with
$C_{22i} = 0$ where fortuitous initial conditions lead to all the
resonance variables at the 6:1 commensurability librating about
$0^\circ$, with the orbits being aligned ($\varpi-\varpi_{\s C} =
0^\circ$).
Unlike the evolution shown in Figs.~\ref{egdtleft} and \ref{egqleft},
this example evolves with $e$ decreasing to nearly zero with
decreasing $e_{\s C}$ as the dual synchronous rotation of
Pluto-Charon is approached. This result might
have been encouraging for the resonant transport of Pluto's small
satellites to their current positions.
However, all of the 6:1 captures were for anti-aligned orbits except
this one, which always lead to increase in $e$ at the end of the tidal
evolution of Pluto-Charon, and we could never capture a test particle
into multiple resonances at the 3:1 and 4:1 commensurabilities.
Finally, our assumption of $J_{2{\s P}}=0$ allowing similar rates of
periapse precession for Charon and the test particle cannot prevail.
The effect of non-zero $J_{2{\s P}}$ is explored in the next section.

\section{EFFECT OF $J_{2{\s P}}$}

After the impact that puts Charon in orbit around Pluto, Pluto would have
absorbed a considerable fraction of the energy dissipated in the
collision. The rapid rotation of a softened Pluto would lead to a
nearly hydrostatic value of the zonal gravitational harmonic $J_{2{\s P}}$.
For rotation about the axis of maximum moment of inertia,
the changes in the principal components of Pluto's moment of
inertia tensor ($\mathcal{A} \le \mathcal{B} \le \mathcal{C}$) from
hydrostatic rotational distortion are (e.g., \citealp{Peale73})
\begin{eqnarray}
\Delta\mathcal{A} = \Delta\mathcal{B}
&=& - {k_{fP} R_{\s P}^5 {\dot\psi}_{\s P}^2\over 9 G} ,
\nonumber \\
\Delta\mathcal{C} &=& + {2 k_{fP} R_{\s P}^5 {\dot\psi}_{\s P}^2\over 9 G} ,
\label{eq:deltaI}
\end{eqnarray}
where $k_{fP}$ is the second degree fluid Love number.
Then
\begin{equation}
J_{2{\s P}}
= \frac{\Delta \mathcal{C} - (\Delta \mathcal{A} + \Delta\mathcal{B})/2}
  {M_{\s P}R_{\s P}^2}
= \frac{k_{fP} R_{\s P}^3\dot\psi_{\s P}^2}{3 G M_{\s P}} ,
\label{eq:j2}
\end{equation}
where we have neglected the tidal contribution to $J_{2{\s P}}$ and any
permanent deviation from axial symmetry. For $a_{\s C}=4R_{\s P}$, the
initial value of $\dot\psi_{\s P}\approx 2\pi/(3.25$ hours), depending on the
initial $e_{\s C}$, leading to $J_{2{\s P}} \approx 0.17$--$0.26$ if
$k_{fP} \approx 1$ (by analogy with the Earth) to $3/2$ (for
homogeneous sphere).
The neglected tidal
contribution to $J_{2{\s P}}$ is approximately 1\% of this value.
We also neglect the contributions of $J_{4{\s P}}$ and $J_{2{\s P}}^2$
to the evolution, since these contributions will only be significant when
Charon is close to Pluto and they will only enhance the effect of
$J_{2{\s P}}$ on the orbital precessions.

We have performed the non-zero $J_{2{\s P}}$ counterparts of the test
particle integrations presented in Section 2. An initial
$J_{2{\s P}}=0.1$ is used, which decreases with $\dot\psi_{\s P}^2$
(see Eq.~[\ref{eq:j2}]),
and the smaller effect of Charon's $J_2$ is ignored.
The initial $J_{2{\s P}}$ is smaller than the above estimate by a
factor of $\sim 2$ but large enough to demonstrate the effect of large
$J_{2{\s P}}$.
Only a few test particles survive the first stage out of all the
integrations in the two tidal models and none of them can be migrated
to the end of the tidal evolution of Pluto-Charon.

The periapse precession rate (to the lowest order) of a satellite of
Pluto is given by \citep{Murray99}
\begin{equation}
\dot\varpi=\frac{3}{2}J_{2{\s P}}n\left(\frac{R_{\s
P}}{a}\right)^2+2\alpha C_1(\alpha)n\frac{M_{\s C}}{M_{\s P}}
\label{eq:apseprec}
\end{equation}
where the first term on the right hand side is the precession induced
by Pluto's $J_2$ and the second term is the additional secular
precession induced in the small satellite's orbit due to Charon.
In Eq.~(\ref{eq:apseprec}), $n = [G (M_{\s P} + M_{\s C})/a^3]^{1/2}$ is the
orbital mean motion of the satellite whose precession is being
determined, $\alpha = a_{\s C}/a$, and $C_1(\alpha)=[2\alpha
(d/d\alpha)+\alpha^2(d^2/d\alpha^2)]b_{1/2}^{(0)}/8$, where
$b_{1/2}^{(0)}(\alpha)$ is the Laplace coefficient.

\begin{figure}[t]
\centering
\epsscale{0.5}
\plotone{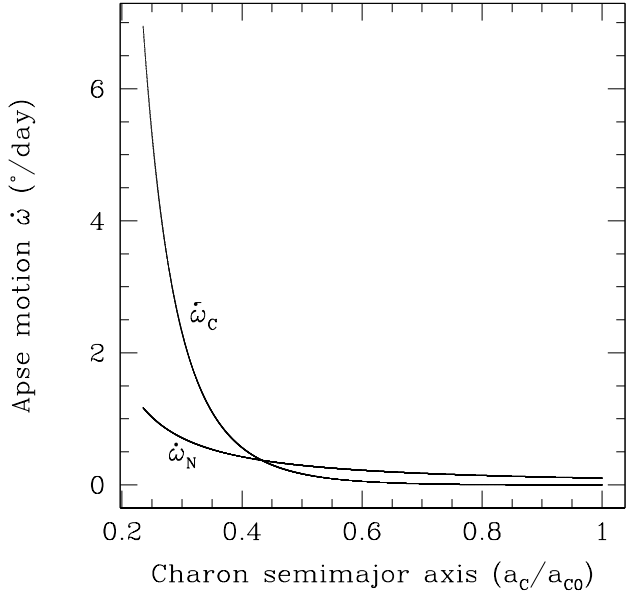}
\caption{
Periapse precession rates of Charon and Nix for hydrostatic value of
$J_{2{\s P}}$ as a function of Charon's semimajor axis.
Nix is assumed to always be at the 4:1 mean-motion commensurability
with Charon.
\label{fig:apseprec}}
\end{figure}

Fig.~\ref{fig:apseprec} illustrates the precession rates of Charon and
Nix as a function of Charon's semimajor axis, where Nix is assumed to
always be at the 4:1 mean-motion commensurability with Charon. Only
the $J_{2{\s P}}$ term is applied to the precession of Charon.
The nominal positions of the resonances can be estimated by
${\dot \phi}_{m,l} \approx 0$ or
\begin{equation}
n_{m,l} \approx \frac{n_{\s C}+l\dot{\varpi}_{\s C}+(m-l)\dot{\varpi}}{m+1}.
\label{nml}
\end{equation}
If $J_{2{\s P}}$ (and hence $\dot{\varpi}_{\s C}$) are assumed to be
zero, the resonances are close to each other, and the $l = m$
corotation resonance is the first one being encountered when the orbit
of Charon expands.
With our estimated initial $J_{2{\s P}} \approx 0.17$ (or larger),
$\dot{\varpi}_{\s C} \gg \dot{\varpi}$ at the 4:1 commensurability and
the resonances are far apart, which means that it is much more
difficult to librate simultaneously in mulitple resonances.
Also, the order of the resonances is reversed and the
corotation resonance is the last one to be encountered.
The resonances exchange their positions when $J_{2{\s P}}$
decreases with $\dot \psi_{\s P}^2$ as Charon moves outward, and any
particles that were captured into multiple resonances become unstable.
Secular precession due to Charon around the 5:1 and 6:1 commensurabilities
should be smaller and the above arguments are applicable.
It is now
clear why the condition of similar precession rates for the
simultaneous capture into and migration within multiple
MMR at the 4:1 to 6:1 commensurabilities cannot be satisfied.
This failure is compounded by the requirement that the
synchronous precession rates must apply simultaneously to all of the
small satellites if they are to be transported within multiple resonances.

\section{INTEGRATIONS WITH MASSIVE NIX AND HYDRA}

Nix and Hydra could be affected by their proximity to the 3:2
mean-motion commensurability if they have non-zero masses.
To see if we overlooked any possible capture into some resonance
configuration that allows Nix and Hydra to be pushed to their current 
distances as Charon's orbit tidally expands, we explore the effects of
non-zero masses for Nix and Hydra.
We ignore Kerberos and Styx in these simulations, as they are likely
much smaller than Nix and Hydra (see footnote \ref{footnote:masses}).

Here we use the Bulirsch-Stoer integrator with the constant $\Delta t$
tidal model described in paper I.
For the constant $\Delta t$ model, the equations of motion in
Cartesian coordinates with the instantaneous tidal forces and torques
and the effects of $J_{2{\s P}}$ and $C_{22i}$ can be integrated
directly (see paper I for details).
As in the previous sections, $k_{2{\s P}} = 0.058$, $\Delta t_{\s P} =
600\,$s, and the initial semimajor axis for Charon's orbit is assumed
to be $a_{\s C} = 4R_{\s P}$ for all of the trials.
We adopt $A_{\Delta t} = 9$.
The initial spin angular velocity of Pluto is determined by the
initial eccentricity of Charon's orbit, consistent with the current
total angular momentum of the Pluto-Charon system.
Charon's spin contributes relatively little to the total angular
momentum, so we arbitrarily choose Charon's initial spin angular
velocity to be half that of Pluto.
The initial angular momenta of Nix and Hydra are neglected in
determining the initial conditions of Pluto and Charon.

\citet{Tholen08} assign albedos of 0.08 and 0.18 to
Nix and Hydra using their best fit masses ($5.8 \pm 5.1 \times
10^{17}\,$kg and $3.2 \pm 6.3 \times 10^{17}\,$kg, respectively) and
assuming Charon's density ($1.63\,{\rm g}\,{\rm cm}^{-3}$) for both
satellites.
The uncertainties in the masses of Nix and Hydra exceed or
are comparable to their best fit values, so we choose to minimize
the masses by assuming that Nix and Hydra have the same albedo as 
Charon of 0.34.
With albedos of 0.34, the radii are reduced by factors of
$\sqrt{0.34/0.08}\approx 2$ and $\sqrt{0.34/0.18}\approx 1.4$,
respectively, thereby reducing the masses by factors of $\sim 8$ and
$\sim 3$.
These reductions in the masses from the \citet{Tholen08}
best fit values lead to $M_{\s N} = 7.25\times 10^{16}\,$kg and
$M_{\s H}=1.1\times 10^{17}\,$kg.
These masses are between the high-albedo masses adopted by
\citet{Lee06} and the upper limits derived by \citet{Youdin12} from
the orbital stability of Kerberos.
By using minimum masses for Nix and Hydra, we maximize the
chances for a stable configuration.

Initial semimajor axes of Nix and Hydra place them outside the 4:1 and
6:1 mean-motion commensurabilities with Charon, and Hydra outside the
3:2 commensurability with Nix ($a_{\s N} > 2.52 a_{\s C}$, $a_{\s H} >
1.31 a_{\s N}$). Eight initial values of $e_{\s C}$ are chosen (0.02
to 0.30 at intervals of 0.04), six values of $a_{\s N}/a_{{\s C}0}$
(0.60 to 0.70 at intervals of 0.02, where $a_{{\s C}0} = 19573\,$km is
the current semimajor axis of Charon), six values of
$a_{\s H}/a_{\s N}$ (1.40, 1.42, 1.44, 1.45, 1.46, and 1.48),
and 36 values of the initial true anomalies for both Nix and Hydra
at intervals of $10^\circ$.  Initial values of $e_{\s N}=e_{\s H}=0.01$.
The initially small values of $e_{\s N}$ and $e_{\s H}$ mean the
periapse positions of both Nix and Hydra will be rapidly scrambled by
the perturbations, so initial values are arbitrary set at
$\varpi_{\s N}=180^\circ$, and $\varpi_{\s H}=0^\circ$, for all the
runs.
Also, the initial value of $\varpi_{\s C}=0^\circ$ for all the runs,
where this longitude will initially precess rapidly ($\sim 5^\circ$/day)
for hydrostatic value of $J_{2{\s P}}$.
The choices of the parameter values lead to
$8\times 6\times 6\times 36\times 36=373,248$ trials for $J_{2{\s P}}=0$
and an equal number of trials for hydrostatic value of $J_{2{\s P}}$
($= 0.17$ initially and decreasing with $\dot\psi_{\s P}^2$; see
Eq.~[\ref{eq:j2}]).
A run is terminated if any eccentricity exceeds 0.7 or any semimajor
axis exceeds $5a_{{\s C}0}$.

There are no survivors for either value of $J_{2{\s P}}$.
The longest time to instability is a little over $10^5$ days, where
longer times to instability corresponded to smaller values of the
initial $e_{\s C}$.
With $J_{2{\s P}}=0$ one could hope that there might be capture of Nix
into multiple resonances with subsequent stable expansion --- at least
until the 6:1 commensurability with Hydra is encountered.
But recall that we get no such captures when Nix is a test particle
(Section \ref{multi}), and that result seems to apply with finite
masses.
Apparently Nix only gets caught into a single MMR, never the
corotation resonance, with subsequent increase in the eccentricity to
the point of instability, as for the case where $J_{2{\s P}}$ is the
hydrostatic value.

\begin{figure}
\centering
\epsscale{1.0}
\plottwo{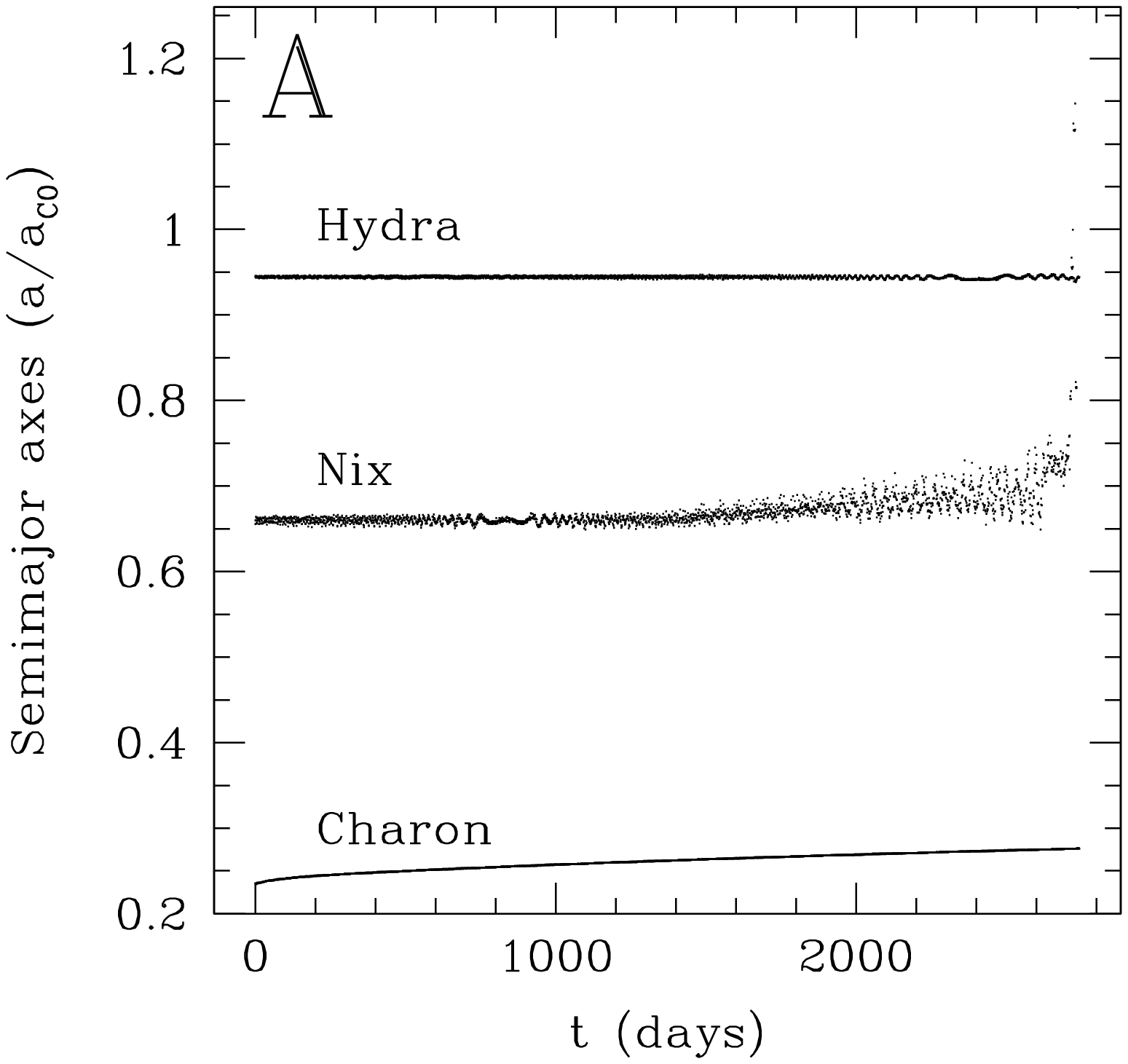}{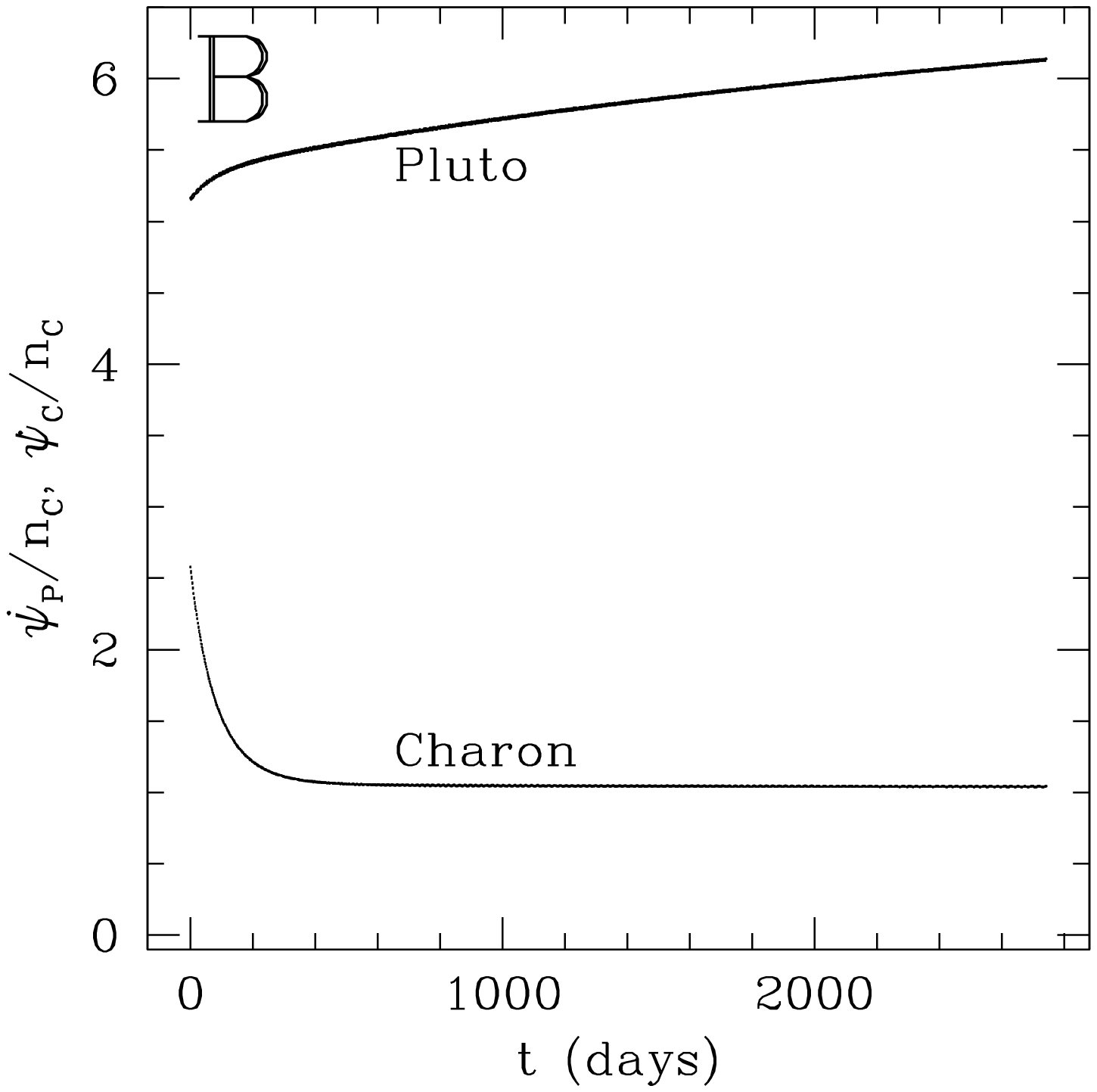}
\vspace{-.5in}
\plottwo{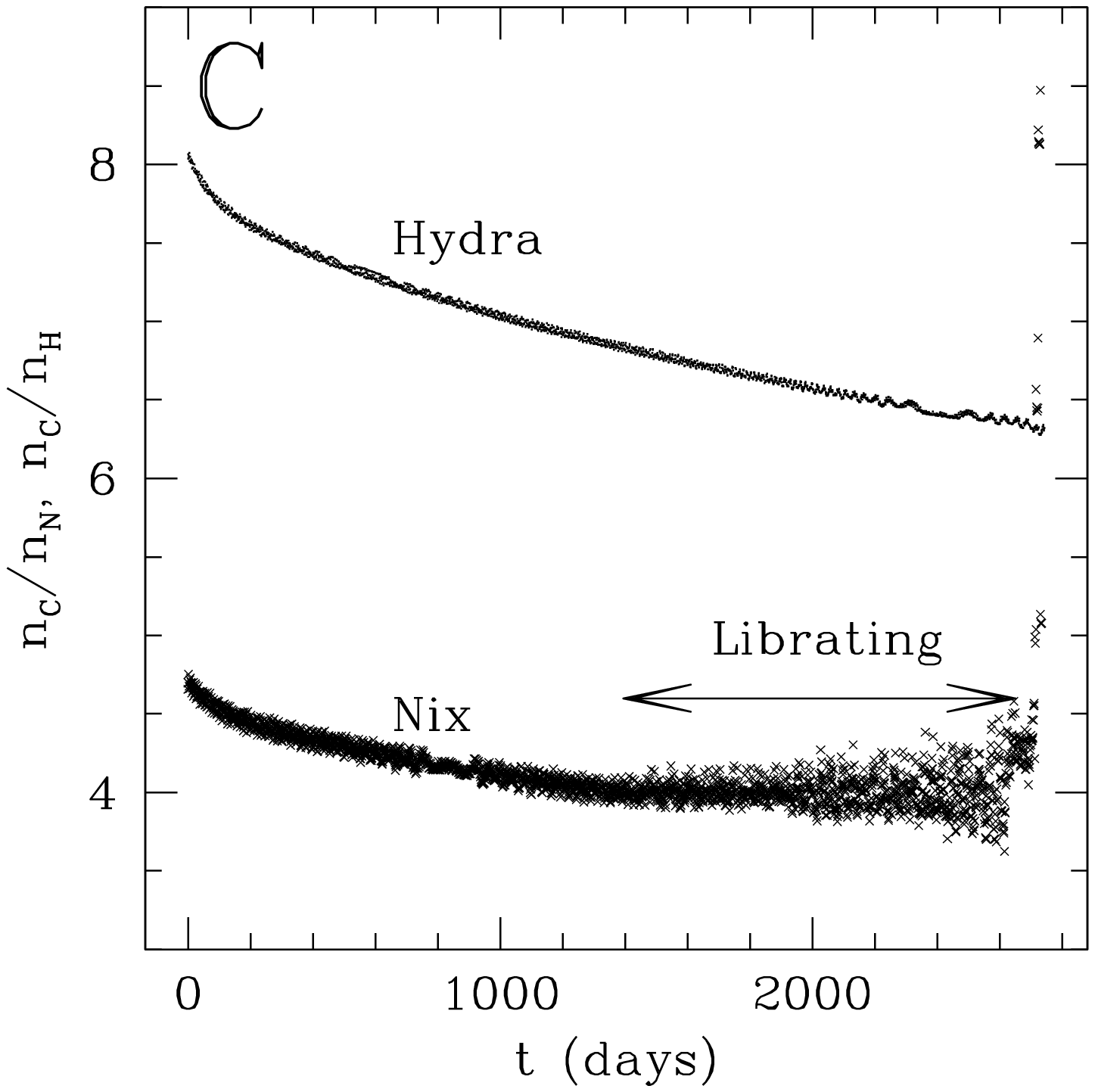}{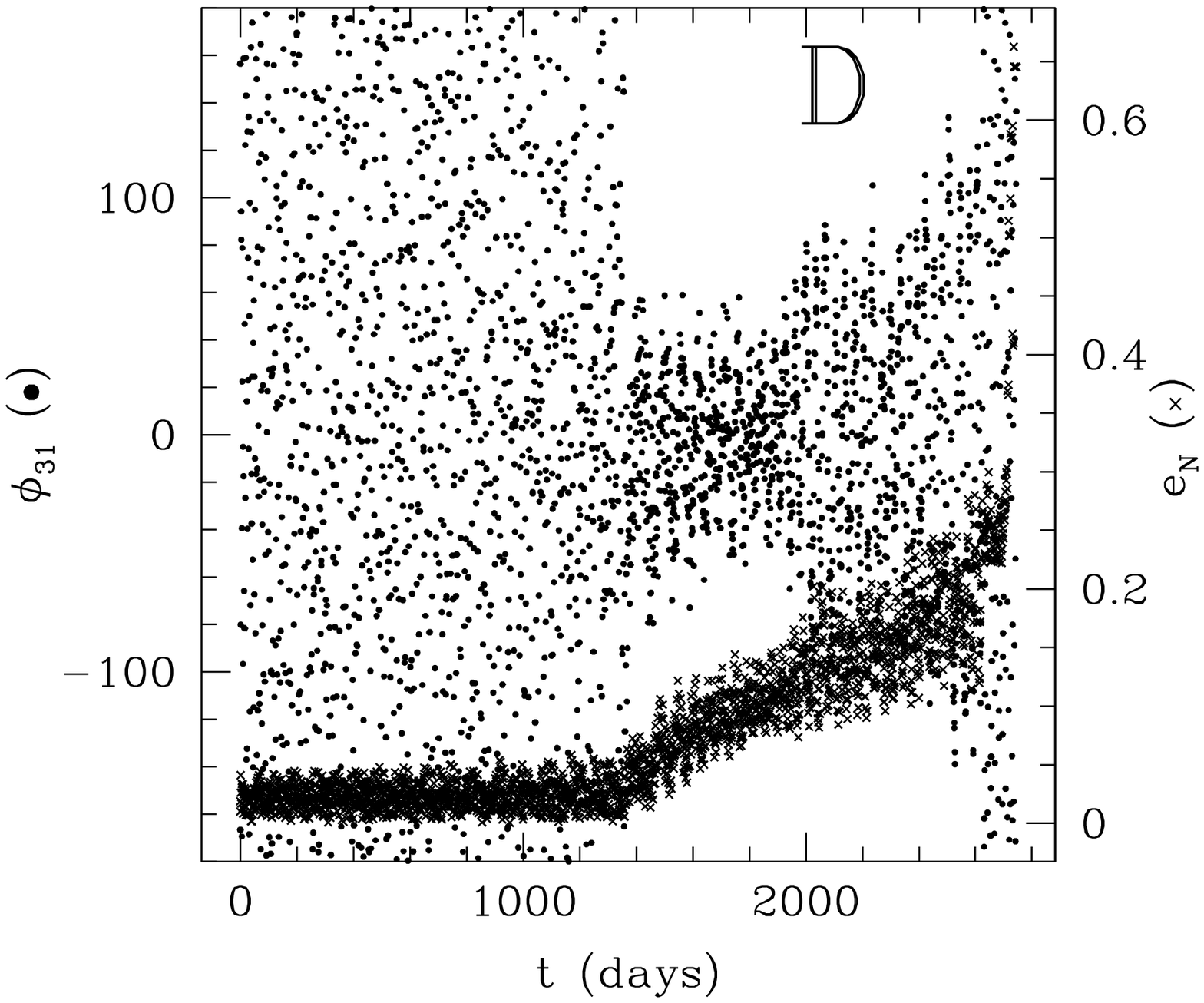}
\caption{
Typical evolution to instability during expansion of Charon's orbit
for simulations with massive Nix and Hydra.
The tidal model is constant $\Delta t$, and $J_{2{\s P}}$ has
hydrostatic value.
Initial conditions: $e_{\s C} = 0.06$, $a_{\s N} =
0.66a_{{\s C}0}$, $a_{\s H} = 1.43a_{\s N}$, $f_{\s N} = 230^\circ$,
$f_{\s H} = 330^\circ$.
\label{fig:realsystem}}
\end{figure}

Fig.~\ref{fig:realsystem} shows the evolution to instability for a typical
integration with hydrostatic value of $J_{2{\s P}}$.
Panel A shows the evolution of the semimajor axes of Charon, Nix and
Hydra, and Panel B shows Charon's and Pluto's spin histories.
Charon's orbit expands as both Pluto's and Charon's spins decrease.
Charon's spin decreases rapidly toward the asymptotic spin state,
which is a little faster than the synchronous spin.
Continued evolution would trap Charon into synchronous rotation
relatively early unless the eccentricity increases.
Nix's semimajor axis is unaffected at first except for short period
fluctuations, but it starts to rise near 1400 days as a result of
being trapped into a 4:1 MMR, where the resonance variable $\phi_{3,1}
= 4\lambda_{\s N}-\lambda_{\s C}-\varpi_{\s C}-2\varpi_{\s N}$ is
librating.
The semimajor axis of Nix's orbit must increase while the resonance
variable is librating in order to preserve the 4:1 commensurability as
Charon's orbit continues to expand.
The value of the ratio $n_{\s C}/n_{\s N}$ stops decreasing at 4:1
near 1400 days as shown in Panel C.
Panel D shows the libration of the resonance variable $\phi_{3,1}$
starting near 1400 days, which libration persists until about 2600
days.
Also shown in Panel D is the rapid increase in Nix's eccentricity as
Nix's orbit is pushed outward by the resonant interaction with Charon.
This increase in $e_{\s N}$ is as expected for evolution within any
single Lindblad resonance (see Section \ref{intro}).
Simultaneous with the increase in $e_{\s N}$, we see in Panel D that
the amplitude of libration increases to the point where the resonance
variable begins circulating, and instability ensues shortly thereafter
with the sudden growth of both $e_{\s N}$ and $a_{\s N}$.

During all of this activity with Nix, Hydra's semimajor axis is almost
undisturbed until oscillations are induced because of the closer
encounters with Nix as the latter's eccentricity grows. This
relatively mild disturbance is shown in both Panels A and C near the
right extremes of the Hydra curves. As the 4:1
mean-motion commensurability between Charon and Nix is always
encountered before the 6:1 commensurability of Charon and Hydra, it is
always Nix that destroys the stability of the system as Charon's orbit
expands.

Since Hydra's orbit is almost undisturbed in the above example, we
repeat the calculation with the same set of parameters but
with Hydra's semimajor axis starting out closer to the 3:2 MMR
with Nix to see if the proximity of this resonance could influence the
outcome of the multi-parameter trial with otherwise the same
parameters. With hydrostatic value of $J_{2{\s P}}$,
there again are no survivors for two runs with the closest initial values
of semimajor axes of $a_{\s H} = 1.34a_{\s N}$ and $1.31a_{\s N}$,
respectively, where the latter value is just outside the 3:2 MMR.  

We are also able to start a calculation with Hydra and Nix in a 
single MMR at the 3:2 mean-motion commensurability with
$3\lambda_{\s H} - 2\lambda_{\s N} - \varpi_{\s H}$ librating about
$180^\circ$ with
amplitude $\sim 20^\circ$.  The latter is established by applying an
artificial drag opposing the velocity of Hydra to ease it into the 3:2
MMR with Nix, with the tidal expansion of Charon's orbit turned off.
We could only get capture into this particular resonance at
the 3:2 commensurability. The purpose of this exercise is to see if
the pre-existence of a Nix-Hydra 3:2 MMR could lead to a stable
expansion of the orbits after all. With initial conditions in the above 3:2
resonance, the tidal expansion of Charon's orbit leads to instability
rather quickly, but not by Hydra's eccentricity increasing because it
is in the single 3:2 MMR, but by Charon's perturbations of Nix
as before --- almost as if Hydra were not there.  Nix gets briefly into a
single 4:1 Lindblad resonance with Charon, with rapidly increasing eccentricity
and ultimate instability. So whether or not Hydra is in a 3:2
resonance with Nix, it is Charon's  perturbations of Nix that lead to the
instability.

Finally, although $\Delta t_{\s P} = 600\,$s is similar to that for
the Earth (see paper I), there is some risk that artifacts can be
introduced if $\Delta t_{\s P}$ is too large and the rate of evolution
is too fast.
Interestingly, calculations with the same initial conditions as those for
Fig.~\ref{fig:realsystem} above, but with $\Delta t_{\s P}$ decreased
by one, two and three orders of magnitude, produce the same overall
evolution with capture of Nix into the 4:1 MMR described above at
about the same value of Charon's semimajor axis $a_{\s C}$.
However, instability ensues at progressively smaller values of
$a_{\s C}$ as $\Delta t_{\s P}$ is decreased, with instability
occurring just as Charon reached the 4:1 commensurability with Nix
when $\Delta t_{\s P} = 0.6\,$s.
The perturbations by the massive Charon accumulate as Charon
spends longer times near a given semimajor axis, leading to the
``earlier'' instability.

\section{CONCLUSIONS AND DISCUSSION}

We have found what we think is the first demonstration of the stable
expansion of a test particle's orbit captured into multiple resonances
at the same mean-motion commensurability with an inner satellite
(i.e., Charon), whose orbit is expanding.
The test particle's orbital eccentricity $e$ does not grow
excessively, as occurs when captured only into a single Lindblad
resonance containing $e$ in the coefficient of the restoring torque.
The hope that such captures would allow the resonant transport of the
small satellites of Pluto and thereby avoid the problem of unlikely
capture into the corotation resonances only (where $e$ would not grow
with continued evolution) was shattered by several observations:
(1) While we could stably migrate a test particle at the 5:1, 6:1 and
7:1 commensurabilities in multiple resonances, we could never stably
migrate a test particle at the 3:1 and 4:1 commensurabilities.
(2) For the test particles stably captured into an orbital
configuration anti-aligned with Charon, final $e$ is too large as $e$
increases with decreasing $e_{\s C}$ on the approach of Pluto-Charon
to the dual synchronous state.
(3) There is only one fortuitous selection of initial conditions for
the constant $Q$ model that leads to the test particle's eccentricity
damping to zero with $e_{\s C}$.
This test particle is captured into an aligned configuration at the 6:1
commensurability.
Our results suggest that the eccentricity evolution in multiple
resonances is related to the geometry of the resonances, with the
evolution following that of Charon in the aligned configuration and
opposite evolution for the anti-aligned configuration, but a better
theoretical understanding of the evolution of test particles in
multiple resonances is needed.
(4) The requirement that Pluto's $J_{2{\s P}} = 0$ for
$\dot\varpi_{\s C} \approx \dot\varpi$, for either anti-aligned or
aligned orbits, does not prevail.
The differential precession of the longitude of periapse of a test
particle and that of Charon for a hydrostatic value of $J_{2{\s P}}$
precludes capture into multiple resonances at the same
commensurability.

To check the parameter space for possible oversight of a stable,
migrating configuration, we integrated the Pluto-Charon system with
finite but minimal masses of Nix and Hydra for a wide range of
plausible initial parameter values that would allow encounter of the
4:1 and 6:1 mean-motion commensurabilities between Charon and Nix and
Charon and Hydra, respectively.
Out of more than $3.7\times 10^5$ trials for $J_{2{\s P}}=0$ and an
equal number of trials for hydrostatic value of $J_{2{\s P}}$,
none of the systems survives.
Placing Hydra closer to or even in the 3:2 resonance with Nix, or
increasing the timescale for the expansion of Charon's orbit, did not
help.
So we conclude that the transport of the small satellites in MMR to
their current distances from Pluto as Charon's orbit tidally expanded
is not possible.

Is there an alternative origin?
\citet{Lithwick08a} have proposed
the creation of a debris disk close to the current locations of the
small satellites after the tidal evolution of Pluto-Charon was
complete.
Such a debris disk, if sufficiently collisional, would settle down
into the plane of the orbit of Charon, where it could accrete into
larger bodies and/or sort the existing debris into long-term stable
orbits.
\cite{PiresdosSantos12} have shown that some planetesimals could be
temporarily captured from heliocentric orbits into orbits around
Pluto-Charon due to the binary nature of Pluto-Charon, with the
capture lifetime $\sim 100$ years.
A debris disk could be formed if the temporarily captured
planetesimals collided with other planetesimals on heliocentric
orbits.
However, for planetesimals large enough to have sufficient mass to
form the small satellites, \cite{PiresdosSantos12} estimated that a
collision during temporary capture is extremely unlikely, because the
timescale for collision is many orders of magnitudes longer than the
capture lifetime.

Alternatively,
\cite{Kenyon14} have suggested that the small satellites grew close
to their current locations from debris ejected by the Charon-forming
impact {\it before} any significant tidal expansion of Charon's orbit.
The debris from the impact was initially located at distances less than $\sim
30 R_{\s P}$ \citep{Canup11}, but \cite{Kenyon14} argued that the
debris would first evolve into a ring at $\sim 20 R_{\s P}$ and then
spread out into a disk out to $\sim 60 R_{\s P}$.
However, if the small satellites formed this way, they would likely
suffer the same problem seen in Section \ref{multi} with the
subsequent tidal expansion of Charon's orbit.
For example, if a small satellite formed near the current location of
Nix at $\approx 43 R_{\s P}$, it would encounter the 7:1 to 5:1
mean-motion commensurabilities with Charon when $a_{\s C} \approx
11.8$--$14.7 R_{\s C}$ or $a_{\s C}/a_{{\s C}0} \approx 0.69$--$0.86$.
For Charon at these distances, the effect of $J_{2{\s P}}$ on its
precession rate is likely small (see Fig.~\ref{fig:apseprec}).
Therefore, when the small satellite encountered one of these
commensurabilities, its orbit would evolve as shown in Section
\ref{multi} and would become either unstable or too eccentric after
stable multi-resonance capture and transport.
It is unclear whether these problems could be overcome if the ratio of
tidal dissipation ($A_{\Delta t}$ or $A_Q$) was much larger than what
we have assumed and Charon's orbit circularized rapidly or if the
orbits of the small satellites could be circularized by interactions
with any remaining debris.

Since none of the proposed scenarios appears viable yet, the origin of
the small satellites of Pluto remains a mystery and further
investigations are needed.
The upcoming {\it New Horizons} flyby may reveal new clues and
constraints.

\acknowledgments
The authors are grateful for the support of a Postgraduate Studentship
at the University of Hong Kong (WHC), Hong Kong RGC Grant HKU 7024/08P
(WHC and MHL), and the NASA Planetary Geology and Geophysics Program
under Grant NNX08AL76G (SJP).
They appreciated useful discussions with Robin Canup, Yoram Lithwick,
and Yanqin Wu.

\newpage

\bibliography{pluto}

\end{document}